\documentclass[aps,pra,twocolumn,superscriptaddress]{revtex4-2}

\usepackage{amsmath,amsfonts,amssymb}
\usepackage{graphicx,float,calc}
\usepackage{color,bm}
\usepackage{ulem}
\usepackage{braket}
\usepackage[colorlinks,urlcolor=blue,citecolor=blue,linkcolor=blue]{hyperref}
\usepackage{array}
\usepackage{multirow}

\begin{document}

\title{Recent progress on disorder-induced topological phases}

\author{Dan-Wei Zhang}\email{danweizhang@m.scnu.edu.cn}
\affiliation{Key Laboratory of Atomic and Subatomic Structure and Quantum Control (Ministry of Education), Guangdong Basic Research Center of Excellence for Structure and Fundamental Interactions of Matter, South China Normal University, Guangzhou 510006, China}
\affiliation{Guangdong-Hong Kong Joint Laboratory of Quantum Matter, Frontier Research Institute for Physics, School of Physics, South China Normal University, Guangzhou 510006, China}
\affiliation{Quantum Science Center of Guangdong-Hong Kong-Macao Greater Bay Area (Guangdong), Shenzhen 518045, China}

\author{Ling-Zhi Tang}
\email{tanglingzhi@quantumsc.cn}
\affiliation{Quantum Science Center of Guangdong-Hong Kong-Macao Greater Bay Area (Guangdong), Shenzhen 518045, China}

\begin{abstract}
Topological states of matter in disordered systems without translation symmetry have attracted great interest in recent years. These states with topological characters are not only robust against certain disorders, but also can be counterintuitively induced by disorders from a topologically trivial phase in the clean limit. In this review, we summarize the current theoretical and experimental progress on disorder-induced topological phases in both condensed-matter and artificial systems. We first introduce the topological Anderson insulators (TAIs) induced by random disorders and their topological characterizations and experimental realizations. We then discuss various extensions of TAIs with unique localization phenomena in quasiperiodic and non-Hermitian systems. 
We also review the theoretical and experimental studies on the disorder-induced topology in dynamical and many-body systems, including topological Anderson-Thouless pumps, disordered correlated topological insulators and average-symmetry protected topological orders acting as interacting TAI phases.
Finally, we conclude the review by highlighting potential directions for future explorations.

\end{abstract}

\date{\today}

\maketitle

\tableofcontents

\section{Introduction}\label{Sec1}

Topological phases of matter have emerged as a central theme in condensed matter physics in the last decades \cite{Hasan2010a,XLQi2011a,Bansil2016a,Chiu2016a,Armitage2018a,YBYang2024a}, since the discovery of integer and fractional quantum Hall effects \cite{Klitzing1980a,Tsui1982a,Thouless1982a}. Subsequent discoveries including the quantum anomalous Hall effect \cite{Haldane1988a}, quantum spin Hall effect \cite{Bernevig2006a}, topological insulators \cite{Hasan2010a,XLQi2011a}, and topological semimetals \cite{Armitage2018a} have unveiled numerous novel topological phenomena and materials, offering a platform to study phases and phase transitions beyond the Landau–Ginzburg paradigm. A topological phase is generally characterized by a topological invariant of bulk states and protected boundary states, which are robust against certain types of weak disorders, such as unidirectional and non-dissipative edge transport in topological band insulators. For sufficiently strong disorders, however, the systems usually become trivial Anderson insulators as the band gaps close and bulk states
are localized due to Anderson localization \cite{Anderson1958a,Evers2008a}. In 2009, a counterintuitive topological insulator driven by disorders from a trivial phase in the clean limit was theoretically discovered, termed as the topological Anderson insulator (TAI) \cite{JLi2009a}. The disorder-induced TAI phase was firstly predicted in 2D quantum spin Hall systems, and has been extended to a wide range of disordered systems.

On the other hand, recent advancements have sought to generalize topological phases from condensed matter materials to various artificial systems, including ultracold atoms \cite{Goldman2016a,DWZhang2018a,Cooper2019a}, photonics \cite{LLu2014a,Ozawa2019a,Haldane2008a}, and superconducting quantum simulators \cite{Schroer2014a,Roushan2014a,XSTan2018a,XSTan2019a,XSTan2021a,JDeng2022a}. 
These artificial systems provide convenient ways to control various factors such as the band structures, dimensions, disorders, and interactions more easily than in real materials. These advantages make them promising platforms for exploring the disorder-induced topological phases and phase transitions in the single-particle and many-body levels. In addition, these  systems can be dynamically engineered, and thus are suitable for investigating   disorder-induced topology in dynamical and even non-equilibrium regimes.  
Moreover, in recent years, the study of topological physics has expanded into non-Hermitian systems \cite{Ashida2020a,Bergholtz2021a,Ghatak2019a,Okuma2023a,Banerjee2023a,HZLi2025a}.
The union of non-Hermiticity \cite{Bender1998a,LFeng2017a,ElGanainy2018a,Miri2019a} and topology concepts 
has resulted in a wide variety of exotic topological phenomena being discovered \cite{Lee2016a,SYYao2018a,FSong2019a,ZGong2018a,HShen2018a,YXu2017a,LJin2019a,DWZhang2020a,DWZhang2020b}. In the presence of disorders, non-Hermitian systems can also exhibit unique localization phenomena with point-gap topology \cite{Hatano1996a,Hatano1997a,Longhi2019a,HJiang2019a,Hamazaki2019a,LJZhai2020a,LZTang2021a,LFZhang2021,LJZhai2022a,JLDong2025a,JFRen2024a,LWang2024a,SZLi2024a,RJChen2025a}.
The required non-Hermiticity, such as gain-and-loss and nonreciprocal couplings, has been engineered in artificial systems in recent years.

There are numerous excellent reviews on complementary aspects of topological phases \cite{Hasan2010a,XLQi2011a,Bansil2016a,Chiu2016a,Armitage2018a,YBYang2024a,Goldman2016a,DWZhang2018a,Cooper2019a,LLu2014a,Ozawa2019a,Ashida2020a,Bergholtz2021a,Ghatak2019a,Banerjee2023a,Okuma2023a,HZLi2025a}. Most of them primarily concentrate on clean systems or topological robustness against weak disorders. This review aims to provide a comprehensive overview of recent advances on disorder-induced topological phases in both condensed-matter and artificial systems. We begin with a brief introduction of real-space topological invariants for disordered systems, followed by a discussion of TAIs induced by random disorders and their experimental realizations. We then introduce theoretical and experimental studies of TAIs with various localization properties of bulk states in quasiperiodic and other aperiodic systems. We next present detailed discussion of the non-Hermitian extensions of TAIs. We focus on unique line-gap and point-gap topologies induced by the combination of disorders and non-Hermiticities, as well as their realizations in artificial non-Hermitian systems. We subsequently illustrate the disorder-induced topology in dynamical systems by highlighting counterintuitive (adiabatic and non-adiabatic) topological pumps \cite{Thouless1983a}. We further survey theoretical and experimental explorations of disorder-induced interacting topological phases in many-body fermionic and bosonic systems. Finally, we conclude with an outlook on some possible directions in this rapidly evolving field.

\section{TAIs induced by random disorders} \label{sec2}

In this section, we will introduce some real-space topological markers for characterizing topological phases in disordered systems, and review the theoretical studies of TAIs induced by random disorders and their experimental realizations in artificial systems and real materials.

\subsection{Real-space topological markers}

Before reviewing disorder-induced topological phases, we briefly introduce some topological markers for noninteracting disordered systems. Disorder breaks the translational symmetry of a lattice system, where the crystal momentum is no longer a good quantum number. Consequently, topological bands and topological invariants defined in momentum space become inadequate. The real-space formulation is required to characterize the topological properties of disordered systems. Here, we introduce real-space constructions of several topological invariants: the winding number for 1D systems in the chiral-symmetric class \cite{MondragonShem2014a,JSong2014a,Altland2014a,Altland2015a,HCHsu2020a,HZhang2023a,Velury2021}, the Chern number \cite{Thouless1982a} and Bott index \cite{Bellissard1994a,Collet2011a,HQHuang2018a,HQHuang2018b} for 2D Chern insulators, and the quadrupole moment for chiral-symmetric quadrupole insulators \cite{Kang2019a,Wheeler2019a,Roy2019a}. These real-space topological markers are well quantized when averaging over disorder configurations of random or quasiperiodic types (and so on), which
provide a tool to identify topological phases in generic disordered systems.

\subsubsection{Winding number}
A prototypical tight-binding Hamiltonian of 1D chiral-symmetric AIII class reads \cite{MondragonShem2014a}
\begin{equation}
	H=\sum_{n}\left\{\frac{g}{2}\left[c_n^\dagger(\sigma_x+i\sigma_y)c_{n+1}+\mathrm{H.c.}\right]+m c_n^\dagger\sigma_yc_n\right\},
\end{equation}
where $\sigma_{x,y}$ denotes Pauli matrices and $c_{n,A}^\dagger$ ($c_{n,B}^\dagger$) denotes creation operator for a particle on sublattice $A$ ($B$) at $n$-th unit cell. Here $m$ and $g$ denote the inter-cell and intra-cell hopping strengths, respectively. This lattice model has the chiral (sublattice) symmetry $\Gamma H\Gamma ^{-1}=-H$, with the chiral symmetry operator $\Gamma =\sum_nc_n^\dagger\sigma_z c_n$. In the absence of disorders and under periodic boundary conditions, one has the Bloch Hamiltonian
\begin{equation} H(k)=\begin{pmatrix}0&h_k\\h_k^*&0\end{pmatrix},
\end{equation}
where $h_k=e^{ik}-im$ and $H(k)$ satisfies the chiral symmetry $\sigma_zH(k)\sigma_z^{-1}=-H(k)$. The topological invariant of the model can be obtained from the winding number of off-diagonal part of $h_k$: 
\begin{equation}
	\nu=\int_{0}^{2\pi}\frac{\partial_{k}h_k}{h_k}\frac{dk}{2\pi i}=\left\{\begin{array}{l}1\mathrm{~~if~~}m\in(-1,1),\\0\text{ ~otherwise.}\end{array}\right.
\end{equation}
Note that $\nu$ can also be expressed as $\nu=\frac{1}{\pi}\int_0^{2\pi}dk A(k)$, where $A(k)=i\sum_{\alpha}^{N_\mathrm{occ}}\langle \Gamma u_\alpha(k)|\partial_k|u_\alpha(k)\rangle$, with $|u_{\alpha}(k)\rangle$ being the Bloch function of band $\alpha$.

\begin{table*}[] 
	\begin{tabular}{|c|c|c|c|c|}
		\hline
		Section & Disorder Type                         & Symmetry Class        & Gap Condition     & Topological Invariant                \\ \hline
		II.B    & \multirow{2}{*}{Random}               & AII                   & Bulk/mobility gap & Quantized conductance                \\ \cline{1-1} \cline{3-5} 
		II.C    &                                       & BDI/AIII              & Bulk/mobility gap & Winding number                       \\ \hline
		III.A   & Quasiperiodic                         & BDI                   & Bulk gap          & Winding number                       \\ \hline
		IV.A    & \multirow{3}{*}{Non-Hermitian random} & Non-Hermitian AIII    & Line gap          & Biorthogonal winding number          \\ \cline{1-1} \cline{3-5} 
		IV.B    &                                       & Non-Hermitian Class A & Line gap          & Biorthogonal Chern number/Bott index \\ \cline{1-1} \cline{3-5} 
		IV.D    &                                       & Non-Hermitian         & Point gap         & Poing-gap winding number             \\ \hline
		V.A-B   & Quasiperiodic                         & A                     & Bulk gap          & Chern number                         \\ \hline
		VI.A    & Random +interaction (fermionic)       & AIII                  & Many-body gap     & Many-body Berry phase                \\ \hline
		VI.B    & Amorphous+interaction (bosonic)       & BDI                   & Many-body gap     & Many-body polarization             \\ \hline
	\end{tabular}
	\caption{ Roadmap of disorder-induced topological phases classified by disorder type, symmetry class, gap condition, and the corresponding real-space topological invariants.}
	\label{roadmap}
\end{table*}

The real-space representation of $\nu$ is well defined in 1D chiral symmetric systems \cite{MondragonShem2014a,JSong2014a,Altland2014a,Altland2015a,HCHsu2020a,HZhang2023a}. For simplicity, we use the equivalent flatband Hamiltonian $Q = P_+ - P_-$, where $P_\pm$ denote projectors into positive and negative part of the energy spectrum, respectively. The chiral symmetry operator $\Gamma $ decomposes as $\Gamma  = \Gamma _+ - \Gamma _-$ with spectral projectors $\Gamma _\pm$. Any chiral-symmetric operator like $Q$ expands as $Q = \Gamma _+ Q \Gamma _- + \Gamma _- Q \Gamma _+$, with off-diagonal blocks satisfying $(\Gamma _\pm Q \Gamma _\mp)^\dagger = \Gamma _\mp Q \Gamma _\pm = (\Gamma _\pm Q \Gamma _\mp)^{-1}$. This yields the covariant off-diagonal terms of the winding number as $Q_{+-} \equiv \Gamma _+ Q \Gamma _-$ and $Q_{-+} \equiv \Gamma _- Q \Gamma _+ = (Q_{+-})^{-1}$. By substituting $\int_0^{2\pi} \frac{dk}{2\pi} \text{Tr}\{\}$ to $\mathcal{T}\{\}$ and $\partial_k \to -i[X, \cdot]$, where $\mathcal{T}\{\}$ denotes trace per volume and $X$ is the position operator, one obtains the real-space winding number \cite{MondragonShem2014a}:
\begin{equation}\label{winding}
\nu=-\mathcal{T}\{Q_{-+}[{X},Q_{+-}]\}.
\end{equation}
{It is worth noting that the real-space winding number remains quantized and nonfluctuating when disorder is turned on, even though the bulk energy spectrum is completely localized. Furthermore, $\nu$ remains robust even after the insulating gap is filled with localized states, provided that the chiral symmetry is preserved in the large-system-size limit. This real-space formulation can also be generalized to characterize $(2n+1)$-dimensional systems with chiral symmetry \cite{MondragonShem2014a}.} 

This formula is valid for 1D disordered systems preserving the chiral (sublattice) symmetry. The corresponding result is self-averaging and independent of the specific disorder realization if the lattice size is sufficiently large. This real-space winding number can be detected by measuring the time evolution of the chiral displacement ${\mathcal{C}}_t = 2\langle \psi(t)| \Gamma X |\psi(t)\rangle$ \cite{Cardano2017a,Manda2023a}, where $|\psi(t)\rangle$ denotes the real-space wave function at time $t$. Its time- and disorder-average $\langle \bar{{\mathcal{C}}} \rangle$ corresponds to the winding number $\nu$ in long-time limit.

\subsubsection{Chern number and Bott index}

The Chern number is a topological invariant for the integer quantum Hall effect or Chern insulators, which is usually computed by integrating the Berry curvature over the 2D Brillouin zone. The Chern number for a generic 2D lattice Hamiltonian projector $P$ relies on its projector properties: idempotence ($P^2=P$) and self-adjointness ($P^\dagger=P$). Furthermore, commutation with translations enables a Bloch-Floquet decomposition \cite{Collet2011a}:
\begin{equation}\label{P}	\langle\bm{x}|P|\bm{y}\rangle=\int_{\mathbb{T}^2}\frac{\mathrm{d}\bm{k}}{(2\pi)^2}e^{i\bm{k}\cdot(\bm{x}-\bm{y})}P_{\bm{k}},
\end{equation}
where $P_{\bm{k}}$ is a projector with rank matching the number of orbitals per primitive cell. Moreover, if the matrix elements $\langle \bm{x}|P|\bm{y}\rangle$  exhibit sufficient decay with $\bm{x}-\bm{y}$, the $P_{\bm{k}}'$ become smooth functions of $k$. Under this condition, the Chern number is well quantized and can be defined as \cite{Avron1989a}
\begin{equation}	{Ch} = 2\pi i\epsilon^{\alpha\beta}\int_{\mathbb{T}^2}\frac{\mathrm{d}\bm{k}}{(2\pi)^2}\operatorname{Tr}\left(P_{\bm{k}}\partial_{k_\alpha}P_{\bm{k}}\partial_{k_\beta}P_{\bm{k}}\right),
\end{equation}
where the symbol "Tr" denotes the trace over local orbital degrees of freedom. Using Eq. (\ref{P}), this expression can be reformulated in real space:
\begin{equation}\label{ChP}
	{Ch} =-2\pi i \epsilon^{\alpha\beta} \mathcal{T} (P[P,X_{\alpha}][P,X_{\beta}]),
\end{equation}
where $\mathcal{T}$ denotes the trace per volume, $\lim_{V\to\infty}\frac{1}{V}\mathrm{Tr}_V\{\cdot\}$ is equivalent to $\int_{\mathbb{T}^2}\frac{\mathrm{d}\bm{k}}{(2\pi)^2}$ in translational-invariant systems without disorders, while $X$ represents the position operator. 
{ Remarkably, Eq.~(\ref{ChP}) maintains stability and quantization even in the strongly disordered systems with energy gap closing in the large-system-size limit, provided that matrix elements $\langle \bm{x}|P|\bm{y}\rangle$ decays rapidly, more precisely if \cite{Prodan2011a}
\begin{equation}
	\sum_{\bm{y} \in \mathbb{Z}^2} |\bm{x}-\bm{y}|^2 |\langle \bm{x}|P|\bm{y}\rangle|^2 \leq \infty
\end{equation}}


Another topological invariant for 2D lattice systems is the Bott index \cite{Bellissard1994a,Collet2011a,HQHuang2018a,HQHuang2018b}, which equals to the Chern number in the thermodynamic limit \cite{Toniolo2022a}. To calculate Bott index, one can first redefine the projector for states below a given Fermi energy $P=\sum_i^{N_{\mathrm{occ}}}|\psi_i\rangle\langle\psi_i|$, where $|\psi_i\rangle$ denotes the $i$-th eigenstate with corresponding eigenvalue
$\epsilon_i$. Then the projected position operators can be obtained as
\begin{equation}
	U=Pe^{i2\pi X}P,~~V=Pe^{i2\pi Y}P,
\end{equation} 
where $X$ and $Y$ denote rescaled coordinates restricted in $[0,1)$. The Bott index, which quantifies the commutativity of $U$ and $V$, is expressed as \cite{Bellissard1994a}
\begin{equation}
	B=\frac1{2\pi}\mathrm{Im}\{\mathrm{Tr}[\log(VUV^\dagger U^\dagger)]\}.
\end{equation}
{ Noted that $B_I$ keeps quantized if $VUV^\dagger U^\dagger$ is non-singular in the large-system-size limit \cite{Loring2010}.}
By further introducing the spin degree of freedom, one can define a spin Bott index to characterize quantum spin Hall states in disordered lattices \cite{HQHuang2018a,HQHuang2018b}.

\subsubsection{Quadrupole moment} 
The quadrupole insulator is a prototypical higher-order topological phase, which is characterized by a quantized bulk quadrupole moment and hosting topologically protected corner modes~\cite{Benalcazar2017a,Benalcazar2017b}. The quadrupole moment maintains quantized value in disordered systems, provided that the chiral symmetry is preserved. This allows us to study disorder-induced higher-order topology in quadrupole insulators. In the real space, the quadrupole moment is given by~\cite{Kang2019a,Wheeler2019a,Roy2019a}
\begin{equation} q_{xy}=\frac{1}{2\pi}\mathrm{Im}\left(\log \left[ \det(U^\dagger Q U)\sqrt{\det(Q^\dagger)} \right]\right).
\end{equation}
Here $Q\equiv\exp[i2\pi XY /(L_xL_y)]$ with $X(Y)$ being the position operator and $L_{x,y}$ the system size, the matrix $U$ is obtained by packing each occupied eigenstate column-wise. 
{ Noted that $q_{xy}$ keeps quantized for chiral-symmetric Hamiltonians in the large-system-size limit\cite{YSHu2021a}.}
In analogy to the real-space winding number, the quadrupole moment can be dynamically extracted via the evolution of the mean chiral quadrupole moment~\cite{Mizoguchi2021a}.

{To provide a comprehensive overview of the diverse physical systems discussed in the following of this review, we summarize the key characteristics of disorder-induced topological phases in Table \ref{roadmap}. This roadmap classifies various phases based on their disorder types, symmetry classes, and gap conditions, while highlighting the specific real-space topological invariants employed to characterize them across different regimes.}

\subsection{Theoretical discovery of TAIs}

\subsubsection{Quantum spin Hall systems}

Topological phases are robust against certain disorders, while they generally become trivial phases under strong disorders. The counterintuitive phenomenon was theoretically discovered in 2D quantum wells with quantum spin Hall states \cite{JLi2009a}. It was found that a trivial insulating or metallic phase in the system can undergo a transition to topologically non-trivial phase upon introducing random disorders. Such a disorder-induced topological insulating phase is termed the TAI. In the seminal work \cite{JLi2009a}, the HgTe/CdTe quantum well in the clean case was considered with the effective Hamiltonian \cite{Bernevig2006a}
\begin{equation} \label{BHZ model}	{H}_{0}(\bm{k})=\begin{pmatrix}h_0(\bm{k})&0\\0&h_0^*(-\bm{k})\end{pmatrix}.
\end{equation}
Here $h_0(\bm{k})={\varepsilon}(\bm{k})+\bm{d}(\bm{k})\cdot\bm{\sigma}$ with the 2D wave vector $\bm{k}=(k_x,k_y)$, Pauli matrices $\bm{\sigma}=(\sigma_x,\sigma_y,\sigma_z)$, $\varepsilon(\bm{k})=C-D(k_{x}^{2}+k_{y}^{2})$, $\bm{d}(\bm{k})=[Ak_{x},Ak_{y},M-B(k_{x}^{2}+k_{y}^{2})]$, and $\{A,B,C,D,M\}$ as sample-specific parameters. When the topological mass $M$ is negative, the system in the clean case hosts a quantum spin Hall state. Conversely, a positive $M$ results in an ordinary insulating phase.

\begin{figure}[t] 
	\centering
	\includegraphics[width=0.48\textwidth]{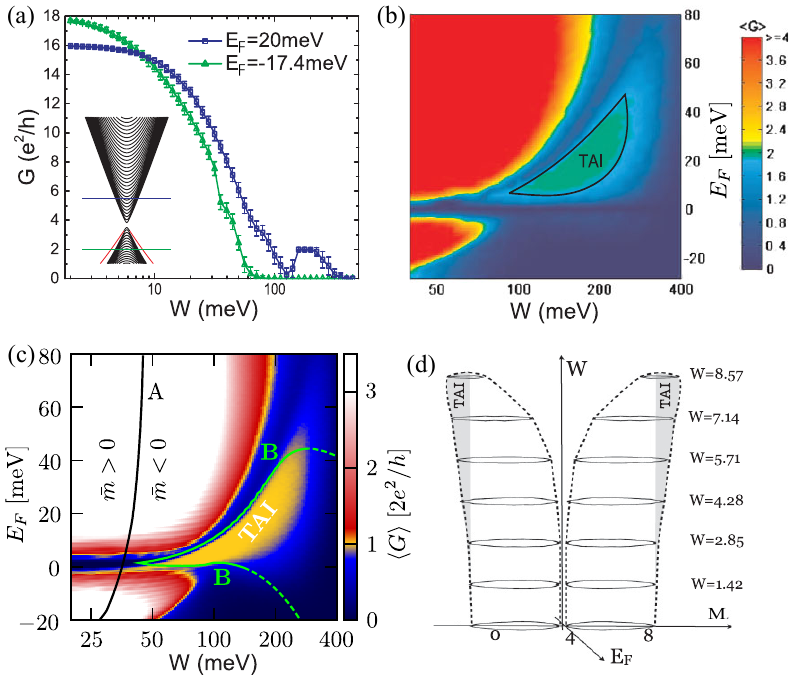}
	\caption{(color online) (a) Conductance $G$ as a function of random disorder strength $W$ for two Fermi energies. The inset show the energy spectrum, where horizontal lines mark Fermi energies. 
	(b) and (c) 2D Phase diagrams showing $\langle G \rangle$ versus $W$ and $E_\mathrm{F}$, with the TAI phase regime. Theoretical phase boundaries from effective medium theory in (c) show as two curves $A$ and $B$ . 
	(a) and (b) are adapted from Ref. \cite{JLi2009a}. (c) is adapted from Ref. \cite{Groth2009a}. (d) 3D phase diagram in parameter space spanned by $E_F$, $W$ and Dirac mass $M$. Contour boundaries combined with dotted line enclose the quantum spin Hall phase regimes, with shaded regime for the TAI. Adapted from Ref. \cite{Prodan2011a}.
	}
	\label{fig1}
\end{figure}

The transport in the corresponding 2D lattice with disorders was numerically studied \cite{JLi2009a}, by introducing random on-site energies with the uniform distribution in the range $[-W/2,W/2]$. Here $W$ denotes the strength of the random disorder. Figure \ref{fig1}(a) displays the two-terminal conductance $G$ for disordered strips versus $W$ at various Fermi energies $E_F$. When $E_F$ resides in the valence band, $G$ rapidly decays to zero with increasing $W$. Conversely, for $E_F$ near the edge of the conduction band, introducing random disorders initially suppresses $G$. Remarkably, with further increasing of the disorder strength, the conductance rebounds and stabilizes at a quantized plateau. This topological plateau of $G=2$ signals the emergence of the topological phase with helical edge states driven from a trivial phase by disorders. These in-gapped helical edge states are absent in the clean limit [the inset of Fig.~\ref{fig1}(a)]. Random disorders in this 2D system localize bulk states and induce helical edge channels, which thereby enable robust quantized transports. The phase diagram in Fig.~\ref{fig1}(b) for $M>0$ shows the $W$-$E_F$ parameter regime for the disorder-induced TAI phase. The detailed edge transport in the TAI phase was sequently studied in Ref. \cite{HJiang2009a}, by numerically calculating the distributions of local currents in the stripe and cylinder samples. It was confirmed that the system becomes the TAI with quantized two-terminal conductance arising from helical edge states at a moderate disorder strength in the two samples.

The physical mechanism for the TAI phase was subsequently elucidated through an effective medium theory and self-consistent Born approximation (SCBA) \cite{Vollhardt1980} in Ref. \cite{Groth2009a}. The random disorder renormalizes the topological mass $M$ and chemical potential $\mu$ as $\bar{M}$ and $\bar{\mu}$, respectively, which can be extracted by the self-energy $\Sigma$ under disorders. For Hamiltonian in Eq. (\ref{BHZ model}), the self-energy is given by 
\begin{equation}
	(E_F-H_0-\Sigma)^{-1}=\langle(E_F-H)^{-1}\rangle,
\end{equation}
where $H$ denotes the tight-binding model obtained by discretizing $H_0$ on a 2D square lattice, $\langle \cdots \rangle$ denotes disorder average. The self-energy $\Sigma$ can be obtained by the SCBA and decomposed as $\Sigma=\Sigma_0\sigma_0+\Sigma_x\sigma_x+\Sigma_y\sigma_y+\Sigma_z\sigma_z$.
The renormalized parameters are given by $\bar{M}=M+\lim_{k\rightarrow0}\mathrm{Re}(\Sigma_z)$ and $\quad\bar{\mu}=E_F-\lim_{k\rightarrow0}\mathrm{Re}(\Sigma_0)$,
with an approximate solution in closed form \cite{Groth2009a}. The phase boundary of the system is determined by $\bar{M}=0$, where the Fermi level $E_F$ enters the inverted gap at $|\bar{\mu}|=-\bar{m}$. The renormalized topological mass $\bar{M}$ and chemical potential $\bar{\mu}$ are obtained as the curve lines A and B in the $E_F$-$W$ parameter plane, as shown in the phase diagram in Fig.~\ref{fig1}(c). In this system, the random disorder induces a negative correction to the topological mass, and makes a negative shift of Fermi level in the conduction band. This explains the emergence of the TAI phase with quantized conductances from a trivial insulating or metallic phase driven by disorders. The phase boundary of the TAI obtained from the effective medium theory is consistent with the numerical results of conductances, but is invalid at strong disorder. Under sufficiently strong disorder, the TAI becomes the trivial Anderson insulator.

The TAI phase in the quantum well system is identical to the quantum spin-Hall phase in a 3D-extended phase diagram \cite{Prodan2011a}, which is formed by Dirac mass $M$, Fermi energy $E_F$, and disorder strength $W$, as shown in Fig.~\ref{fig1}(d). Each 2D phase diagram represents a specific cross-section extracted from the comprehensive 3D phase diagram. Utilizing appropriate parameters for the system, one can observe the distortion of topological phase boundaries with increasing disorder strength. The quantum spin-Hall phase extends to the TAI regime in the 3D parameter space, which belongs to a trivial phase in the clean limit. In this sense, the TAI in the 2D quantum well systems represents a disorder-driven manifestation of the quantum spin-Hall phase \cite{SLZhu2006a}. The TAI phase can be also understood from the disorder-induced band touching~\cite{YYZhang2012a}, originating from either trivial or nontrivial parent bands~\cite{YYZhang2013a}. The associated disorder-induced topological phase transition can be described microscopically as a percolation transition~\cite{Girschik2015a}. The critical exponent of the metal–TAI transition has been estimated, which coincides with that of the symplectic class~\cite{Yamakage2013a}. The stability and phase structure of the TAIs with respect to the finite-size effect~\cite{WLi2011a,DXu2012a} and the effect of spatial correlations of disorders \cite{Girschik2013a,JTSong2012a} have been studied. It was numerically found that there are two distinct scaling regions of the TAIs with and without bulk gaps \cite{DXu2012a}, respectively.

\subsubsection{Other systems}
The TAIs in other 2D randomly disordered systems have also been revealed, such as 2D weak AIII insulator \cite{Claes2020a}, Chern insulator models with quantum anomalous Hall states~\cite{YXXing2011a,Vu2022a,YSu2016a,Sobrosa2024a,Addison2025a}, 
spin-orbit coupled Lieb lattices with both quantum anomalous and spin Hall states \cite{RChen2017a}, (decorated) honeycomb lattices with disordered exchange couplings~\cite{Chua2011a} and disorder-recovered average symmetries \cite{JZhang2022a}, magnetically doped thin films \cite{Okugawa2020a,Okugawa2022a,XXYi2024a}, and under irradiation of circularly polarized light~\cite{ZNing2022a}. Recent advances establish the universality of TAIs in 2D honeycomb lattices~\cite{SLZhu2007a,DWZhang2011} and identify novel topological transitions driven by random flux therein~\cite{Orth2016a,Skipetrov2022a,CALi2025a}. In the presence of
geometric random-bond disorders, a robust metal can emerge in a 2D Chern insulator with particle-hole symmetry \cite{Pachhal2025a}. It has been predicted that disorder from randomly oriented magnetic moments can induce a so-called quantum anomalous parity Hall phase protected by reflection symmetry \cite{Haim2019a}.

The disorder-induced topological phenomena can also be studied in other dimensions. The TAIs have been revealed in the 1D chiral-symmetric AIII and BDI class systems with random disorders ~\cite{MondragonShem2014a,JSong2014a,Altland2014a,Altland2015a,HCHsu2020a,HZhang2023a}, such as the Su-Schrieffer-Heeger (SSH) model \cite{WPSu1979a}. Recently, the 1D TAIs has been extended to flat-band models \cite{ZWZuo2024a} and various types protected by latent symmetry \cite{JRLin2025a}. In 3D systems across various symmetry classes, it has been shown that disorders allow for transitions between metallic, trivial, and TAI phases~\cite{HMGuo2010b,Ryu2012a,Kobayashi2013a}. The 3D antiferromagnetic topological insulator can be induced by disorders from a trivial insulator \cite{Baireuther2014a}. Moreover, two Weyl nodes in Weyl metals can be annihilated pairwise by disorder scattering, leading to a transition from a Weyl metal phase to a 3D quantum anomalous Hall state \cite{CZChen2015a,YWu2017a}. By adding random disorders onto a 3D axion insulator, the bulk and surface topology can be decoupled, which yields a 2D unquantized anomalous Hall metal \cite{Grindall2025a}. The TAI phase has even been theoretically extended to 4D disordered systems, characterized by a second Chern number~\cite{RChen2023a}.

The disorder-induced topology has been explored in the context of superconductors. For instance, in the case of a 2D $d$-wave superconductor, the real-space Chern number reveals a random disorder induced topological phase \cite{Borchmann2016a}. A topologically trivial superconductor can be driven into a chiral topological superconductor upon doping magnetic disorder in 2D spin-orbit coupled superconductors \cite{WQin2016a,SLZhu2011a}. In 1D $p$-wave superconductors, strong disorders usually lead to a transition from a topological superconducting phase to a topologically trivial localized phase \cite{XCai2013a,Degottardi2013a}. However, it was found that in long-range or dimerized Kitaev superconductor chains \cite{Kitaev2001a}, disorders can induce topological phases hosting Majorana zero-energy edge modes  \cite{Lieu2018a,Habibi2018,CHua2019a}.

Furthermore, the disorder-induced topology has recently been extended to bosonic and higher-order topological systems \cite{YBYang2024a}. For instance, the disorder-induced magnonic TAIs in 2D ferromagnets were evidenced from the bosonic Bott index \cite{XSWang2020a}. The higher-order topology is robust against disorders \cite{ZXSu2019a,Araki2029a}, and disorders can even induce quantized multipole moments in higher-order systems, giving rise to the higher-order TAIs \cite{CALi2020a,YBYang2021a,WXZhang2021a,Franca2019a,Ghosh2023a}. Meanwhile, the classical analog of such higher-order TAIs was experimentally constructed using electric circuits and the disorder-induced corner states were observed via the voltage measurement \cite{WXZhang2021a}. In the 2D higher-order TAI systems, the corner states are uniquely protected by a mobility gap rather than the bulk gap \cite{YSHu2021a}. For chiral-symmetric higher-order topological phases under disorders \cite{Benalcazar2022a}, the topological characters can be extracted by the mean chiral quadrupole moment from quench dynamics \cite{Mizoguchi2021a}. Successive disorder-driven phase transitions were revealed in 2D higher-order TAIs \cite{ADLi2025a,YLSong2024a}. The second-order TAI phase was also studied in 2D amorphous and fractal lattices \cite{TPeng2025a,HChen2023a}. In 3D systems, the phase transitions with second- and third-order TAIs induced by disorders have been explored in Refs. \cite{CWang2020a,ZQZhang2021a,JHWang2021a,Hugo2024a,YShen2024a}. We note the review \cite{YBYang2024a} for more discussions of the higher-order topological phases in the absence and presence of disorders.

\subsection{Experimental realizations in artificial systems and real materials}

We now introduce recent experimental realizations of random-disorder-induced TAIs in ultracold atoms, photonic (and other classical wave) systems, and solids.

\subsubsection{Ultracold atoms}

\begin{figure}[t] 
	\centering
	\includegraphics[width=0.48\textwidth]{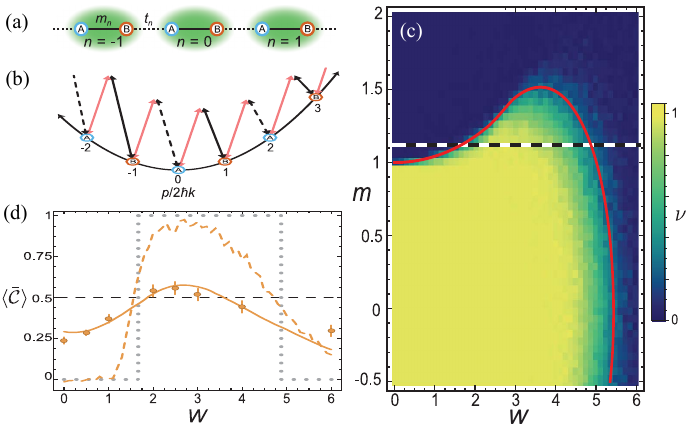}
	\caption{Observation of the TAI in disordered atomic wires.
(a) Schematic lattice of the chiral symmetric wire. (b) Schematic of the experimental implementation of the tight-binding model engineered with atomic momentum states. (c) Topological phase diagram calculated by the real-space winding number $\nu$. (d) Experimental measurement of time- and disorder-averaged chiral displacement $\langle\bar{\mathcal{C}}\rangle$ as a function of disorder strength $W$. Solid and dashed lines are simulations for the same lattice and for larger system, respectively. Dotted gray curve denotes $\nu$  in the thermodynamic limit. Adapted from Ref. \cite{Meier2018a}.}
	\label{fig2}
\end{figure}

The disorder-induced TAI was first experimentally realized with ultracold atoms \cite{Meier2018a}. In the experiment, a 1D chiral symmetric wire with controllable random disorders was synthesized, based on the laser-driven coupling of discrete momentum states of bosonic atoms, as shown in Figs.~\ref{fig2}(a) and (b). The tight-binding model of the 1D chiral symmetric wire consists of two-site unit cells, with sublattice sites A and B. The Hamiltonian reads
\begin{equation}
	H=\sum_n\left\{ m_nc_n^\dagger Sc_n+\frac{t_n}{2}\left[c_{n+1}^\dagger{(\sigma_x-i\sigma_y)}c_n+\mathrm{H.c.}\right]\right\},
\end{equation}
where $c_n^\dagger=(c_{n,\mathrm{A}}^\dagger,c_{n,\mathrm{B}}^\dagger)$ creates a particle at unit cell $n$ in sublattice site A or B, and $\sigma_i$ are Pauli matrices related to the sublattice degree of freedom. The $m_n$ and $t_n$ characterize the intra- and intercell tunneling energies, respectively. This model obeys $\Gamma H\Gamma=-H$ with the chiral operator $\Gamma=\sigma_z\otimes\mathbb{I}$, and can describe chiral wires of the AIII or BDI symmetry classes, by choosing the intracell hopping term to be $S=\sigma_x\left({\mathrm{BDI}}\right)$ or $S=\sigma_y\left({\mathrm{AIII}}\right)$. They represent all possible distinguishable chiral classes for which the real-space winding number is defined as the topological invariant. In the experiment, pure atomic tunneling disorder (without the site-potential disorder) is created to preserve the chiral symmetry.

The controlled fluctuations of hopping terms $t_n=t(1+W_1\omega_n)$ and $m_n=t(m+W_2\omega'_n)$ are experimentally realized. Here $t$ is the characteristic intercell tunneling energy, $m$ is the ratio of intra- to intercell tunneling in the clean limit, $\omega_n$ and $\omega'_n$ are independent random real numbers chosen uniformly from the range [-0.5,0.5], and $W_{1}$ and $W_{2}$ are the dimensionless disorder strengths applied to inter- and intracell tunneling. The topological transition in the disordered BDI-class wire and the disorder-induced TAI in the AIII-class wire are studied \cite{Meier2018a}. Figure \ref{fig2}(b) shows the topological phase diagram of AIII-class wires, which is obtained via the real-space winding number $\nu$ for $W_{2}=W$ and $W_{1}=0$. In the clean limit, the critical point between the topological and trivial phase is at $|m|=1$. The random tunneling disorder induces the TAI phase for $m$ just exceeding 1 over a range of weak to moderate $W$ values. For large disorder, the TAI eventually gives way to a trivial Anderson insulator phase. By measuring the bulk evolution of the atomic gas after sudden quenches, the finite time- and disorder-average chiral displacement $\langle{\mathcal{C}}\rangle$ is extracted \cite{Cardano2017a}, which converges to the winding number $\nu$ in the long-time and thermodynamic limits. Figure \ref{fig2}(c) shows $\langle{\mathcal{C}}\rangle$ as a function of the disorder strength $W$ for an atomic wire of 20 unit cells. For weak disorder, $\langle\bar{{\mathcal{C}}}\rangle$ increases and reaches a maximum at $W \approx 2.5$, before decaying under strong disorder. This initial rise followed by a decrease signals the first transition from a trivial insulator in the clean limit to the TAI phase, and subsequently to a trivial Anderson insulator at strong disorder strength. Notably, the topological phase with critical localization was observed for ultracold atoms in a momentum lattice \cite{TXiao2021a}, which mimics a 1D generalized Aubry-André-Harper model \cite{Aubry1980,Harper1955} with both diagonal and off-diagonal quasi-periodic disorders.

\subsubsection{Photonic systems}\label{SecIIC2}

\begin{figure}[t] 
	\centering
	\includegraphics[width=0.47\textwidth]{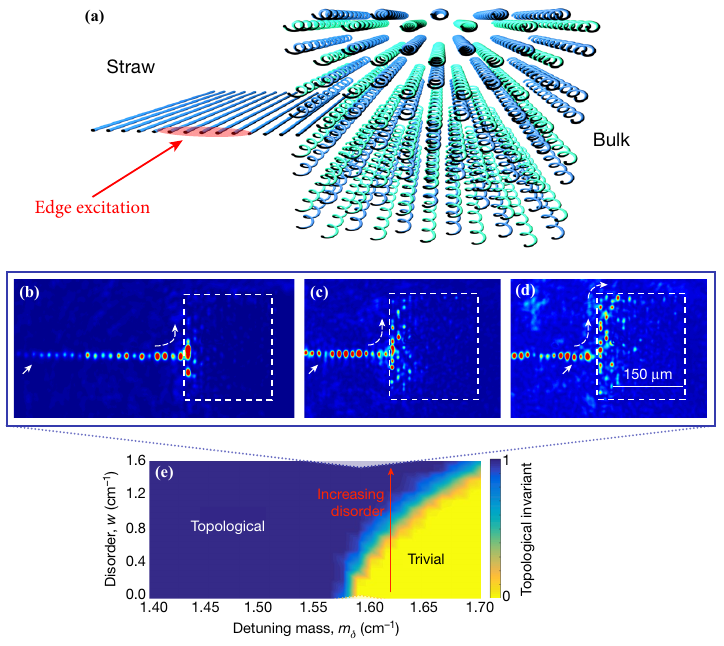}
	\caption{Photonic TAIs. (a) Schematic of the 2D waveguide configuration
and a 1D straw, through which the edge modes are excited. 
(b-d) The formation of chiral edge states when sufficient disorder is added.
(e) Topological Phase diagram showing the real-space Chern number as a function of the detuning mass $m_{\delta}$ and disorder strength $W$. Adapted from Ref. \cite{Stuetzer2018a}.}
	\label{fig3}
\end{figure}

At the same time, a complementary work reported the experimental demonstration of a photonic TAI \cite{Stuetzer2018a}. The experiment was carried out in a photonic platform, based on a 2D Floquet topological insulator with broken time-reversal symmetry, as theoretically proposed in Ref. \cite{Titum2015a}. The platform consists of an array of evanescently coupled helical waveguides, as shown in Fig.~\ref{fig3}(a). Classical light propagation in the system follows the paraxial wave equation, which is mathematically equivalent to the Schrödinger equation for 2D topological insulators. The propagation distance takes the role of time. In the absence of disorders, there are no observable edge excitation and minimal bulk penetration. Introducing sufficient on-site random disorders, the unidirectional edge transport is observed, as shown in Figs.~\ref{fig3}(b-d). The disorder-induced chiral mid-gap edge states demonstrate the formation of the TAI in this 2D photonic system. Figure \ref{fig3}(e) shows a phase diagram that indicates the topological invariant (the real-space Chern number) that counts the number of in-gap edge states as a function of detuning $m_{\delta}$ and disorder strength $W$. The phase diagram shows the TAI phase with chiral edge modes emerges when sufficient on-site random disorder is introduced. The robustness of the topological edge states and disorder-induced topological transitions have been experimentally demonstrated in photonic metamaterials and amorphous lattices \cite{CXLiu2017a,PHZhou2020a}. Notably, the concept of TAIs has been experimentally verified in other classical wave systems, such acoustic waveguides \cite{ZangenehNejad2020a,SFLi2022a,HLiu2023a}, electric circuits \cite{ZQZhang2019a,WXZhang2021a} and mechanical systems \cite{XTShi2021a}.

The recent achievements in engineering disorders in photonic systems enable the exploration of topology of light \cite{SKYu2021a}. A photonic TAI with robust chiral edge states in the microwave regime has been observed in a 2D disordered gyromagnetic crystal \cite{GGLiu2020a}. An experimental scheme has been proposed to realize the photonic $\mathbb{Z}_2$ TAI without pseudospin conservation \cite{XHCui2022a}. Another scheme for realizing the TAI of light in a disordered atomic lattice was proposed in Ref. \cite{Skipetrov2022a}. A photonic TAI with quantum spin-Hall states without breaking time-reversal symmetry was observed in a disordered metacrystal, which is confirmed via the unidirectional propagation and robust transport of helical edge modes \cite{XDChen2024a}. Beyond gapped topological phases, the observation of ungapped TAIs featuring zero-energy bulk degeneracy in coupled photonic resonators has been reported in Ref. \cite{MNRen2024a}. Moreover, a programmable topological photonic chip with large-scale silicon-based integrated circuits has been developed \cite{TXDai2024a}, which enables the comprehensive studies of various topological phenomena, including the counterintuitive TAI phase transition. This photonic chip provides a platform for engineering classical and quantum topological phases induced by disorders. Another photonic simulator based on 1D discrete-time quantum walk of photons has been realized to probe the quantum phase transition between topologically distinct Anderson insulator phases \cite{Barkhofen2024}. A time-staggered profile in the circular photon polarization was observed within the setup, which signals the topological Anderson criticality as suggested in Ref. \cite{Bagrets2021a}. Recently, the concept of topological photonic alloy was proposed and experimentally demonstrated in a disordered 2D photonic crystal, where nontrivial topology can emerge by randomly mixing nonmagnetized and magnetized rods \cite{TTQu2024}. It was further shown that the frequency range of the topological band gap of the photonic alloy can be outside the band-gap frequency range of the two mixing crystals \cite{TTQu2024a}.

\subsubsection{Solid-state materials}
 
The realization of TAIs in solid-state systems is elusive due to challenges in controlling disorder. Recently, a 2D topological insulator with localized bulk states and robust edge currents in a HgTe-based semimetal is experimentally observed \cite{Khudaiberdiev2025a}. In this system, there are edge states shunted by the conducting bulk states in the clean limit. Strong disorder induces a mobility gap in the bulk and thus enables the 1D edge states protected from Anderson localization. Nonlocal transport measurements demonstrate the topological protection of edge states under disorders. This is 
different from the original 2D TAI driven from an trivial phase without topological edge states by increasing disorder \cite{JLi2009a}.

Another recent work reported the experimental observation of a unique variant of the TAI in a monolayer MnBi$_4$Te$_7$ Hall bar device \cite{AQWang2025a}. The system originally hosts a trivial insulating state with Chern number ${Ch} = 0$ in clean limit. Under on-site random magnetic disorders, a Chern insulator state with ${Ch}=1$ signed by the coexistence of a single edge channel with quantized Hall plateau and zero longitudinal resistance was observed, as theoretically predicted in Ref. \cite{CZChen2021a}. The trivial-to-nontrivial transition in this monolayer device can be attributed to disorder, as evidenced by universal conductance fluctuations. Akin to the TAI, such a disorder-induced Chern insulator is termed as topological Anderson Chern insulator \cite{AQWang2025a}. This work substantiates the existence of the long-sought TAIs in real materials, and potentially broadens the parameter space for realizing Chern insulators.

\section{TAIs in quasiperiodic systems}

There exist TAIs with various localization properties of bulk states in quasiperiodic and other aperiodic systems. In this section, we will review these TAI phases and their realizations in superconducting quantum simulators and photonic lattices.

\subsection{Gapped TAIs with different bulk states}

\subsubsection{Model and topological phase diagrams}
Here we consider the tight-binding Hamiltonian of a generalized SSH model with quasiperiodic hopping disorders \cite{LZTang2022a}
\begin{equation}\label{gSSH} 
	H_\text{gSSH}=\sum_{n=1}^N(m_na_n^\dagger b_n+g a_{n+1}^\dagger b_n+\text{H.c.}).
\end{equation}
Here $N$ denotes the number of unit cells, $a_n^\dagger$ ($b_n$) creates (annihilates) a particle on sublattice $A$ ($B$) in cell $n$, while $g$ and $m_n$ represent the uniform inter-cell and space-dependent intra-cell hopping strengths, respectively. We implement quasiperiodic modulation on the intra-cell hopping term as $m_n=m+W\cos(2\pi\alpha n + \theta)$, where $m$ represents the intra-cell hopping amplitude, $W$ indicates the strength of quasiperiodic disorder, $\theta$ denotes additional phase shift, and incommensurate modulation comes from irrational number $\alpha$. When $W=0$, Eq. (\ref{gSSH}) reduces to the clean SSH model, which exhibits a topological phase for $m<g$ and a trivial phase for $m>g$. Here, we set $g\equiv1$ as the energy unit, $\alpha=(\sqrt{5}-1)/2$ (golden ratio), $\theta=0$, and system size $N=L/2=610$ with lattice site $L$. Periodic boundary conditions are used throughout unless specified.

\begin{figure}[tb]
	\centering
	\includegraphics[width=0.45\textwidth]{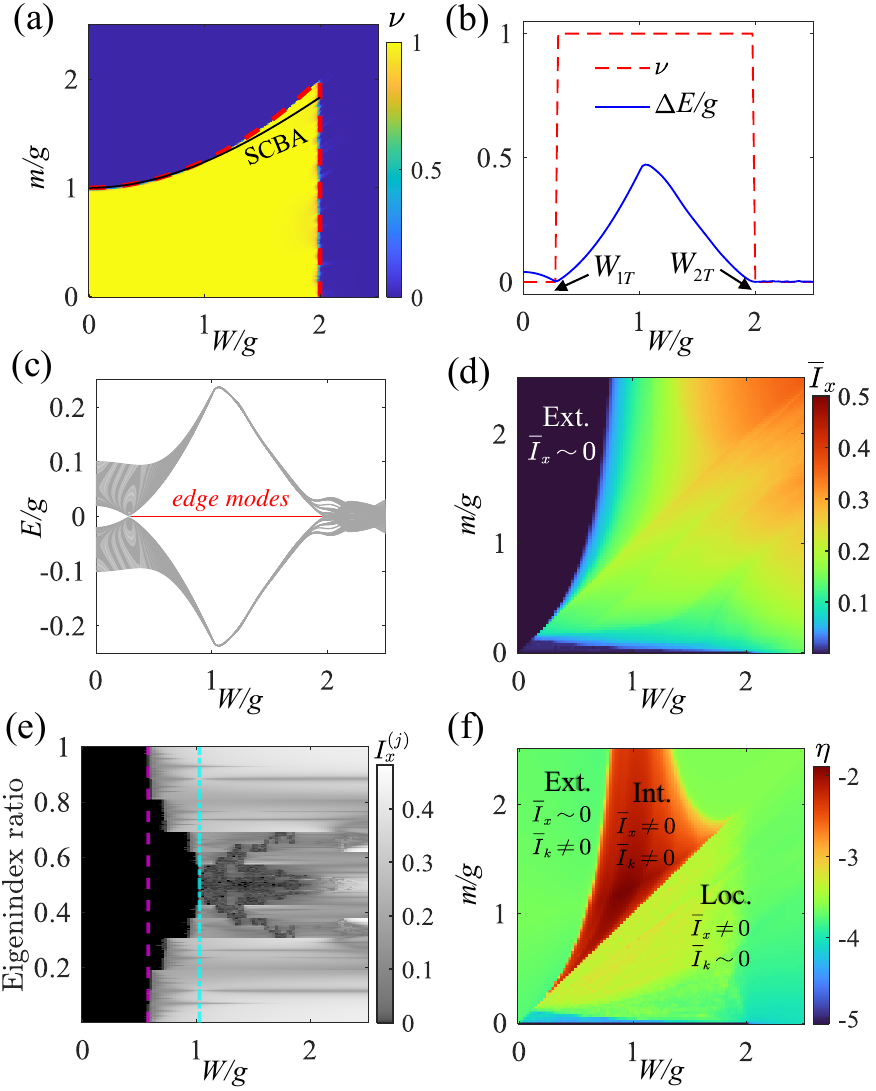}
	\caption{(Color online) (a) Real-space winding number $\nu$ versus $W$ and $m$. Topological phase boundaries (red dashed: zero-energy state localization length divergence; black solid: SCBA analysis) are shown. (b) $\nu$ (red dashed) and $\Delta E$ (blue solid) versus $W$, with topological transitions at $W_{1T}$ and $W_{2T}$. (c) OBC eigenenergies (middle 100 states) versus $W$, highlighting zero-energy edge modes (red) in the topological regime $W_{1T}<W<W_{2T}$. 
    Mean IPRs $\bar{I}_x$ (d) and quantity $\eta$ (f), and IPR $I_{x}^{(j)}$ versus $W$ (e). Regions labeled 'Ext.', 'Loc.', and 'Int.' in (d, f) denote extended, localized, and intermediate phases, respectively. Vertical lines in (e) mark the first ($W_{1L}\approx0.57$) and second ($W_{2L}\approx1.02$) localization transitions.
    {$m=1.02$ is fixed in (b,c,e).}
	Adapted from Ref. \cite{LZTang2022a}. 
	}\label{fig4}
\end{figure}

For the disordered chiral chain, one can use the real-space winding number $\nu$ given by 
Eq. (\ref{winding}) to characterize the topological properties. Figure \ref{fig4}(a) shows the topological phase diagram. At the clean limit, a topological transition occurs at $m=1$ between trivial ($\nu=0$) and topological ($\nu=1$) phases. 
The quasiperiodic disorder can induce the TAI phase with a bulk gap from a trivial phase when $1<m\lesssim2$. For clarity, Fig.~\ref{fig4}(b) shows $\nu$ and $\Delta E$ versus $W$ at fixed $m=1.02$, revealing the disorder-induced gapped TAI between topological transitions $W_{1T} \approx 0.27$ and $W_{2T} \approx 2.0$. The corresponding energy spectra in Fig.~\ref{fig4}(c) under open boundary condition exhibit disorder-induced zero-energy in-gap edge modes in the gapped TAI phase due to the bulk-boundary correspondence.

For zero-energy wave function, the localization length $\Lambda$ diverges at the topological transition points in 1D chiral chain \cite{MondragonShem2014a}. By solving the Schr\"{o}dinger equation for the 
zero-energy wave function, one can derive the expression of $\Lambda$ in the thermodynamic limit and obtain the topological phase boundary where $\Lambda^{-1}\rightarrow0$ \cite{LZTang2022a}, as marked by red dashed lines in Fig.~\ref{fig4}(a). One can also apply the 
SCBA to reveal the disorder effect on the topological phase boundary. The disorder-induced self-energy $\Sigma(W)$ can be obtained by the self-consistent equation:
\begin{equation}
	\frac{1}{E_F-H_q(k)-\Sigma(W)}=\braket{\frac{1}{E_F-H_{\text{eff}}(k,W)}},
\end{equation}
with momentum space Hamiltonian $H_q(k) = [m + g\cos k]\sigma_x + g\sin k\sigma_y$, and self-energy $\Sigma = \Sigma_x \sigma_x + \Sigma_y \sigma_y$.
Here $E_F \equiv 0$ denotes Fermi energy and $\langle \cdots \rangle$ means average over disorder realizations. For the quasiperiodic disorder, the effective Hamiltonian becomes $H_{\text{eff}} = H_q(k) + W\cos(2\pi\alpha n)\sigma_x$ and the self-energy is simplified as $\Sigma(W) = \Sigma_x(W)\sigma_x$. Thus, the intra-cell hopping is renormalized by disorders as $\bar{m} = m + \Sigma_x(W)$, and the topological phase boundary in the $m$-$W$ plane follows $\bar{m}(m,W) = g$. As shown in Fig.~\ref{fig4}(a) (black solid line), the renormalization prediction for the TAI under moderate disorders ($W \lesssim 2$) aligns closely with the numerical results of the winding number.

\subsubsection{Localization properties of bulk states}

The bulk states in this quasiperiodic chiral chain have various localization properties, in contrast to those that are fully localized in 1D random systems. The localization property can be characterized by the inverse participation ratio (IPR) that can be defined in real or momentum space. The IPR of an eigenstate is given by
\begin{equation}\label{IPR}
	I_{\beta}^{(j)}= \sum_{l=1}^{L}\left|\phi_{\beta}^{(j)}(l)\right|^{4},
\end{equation}
where $j$ denotes the eigen-index, $\beta = x, k$ labels the real and momentum space, respectively, Here $\phi_{\beta}^{(j)}(l)$ represents the probability amplitude at site $l$ in $\beta$-space. One can calculate $I_{x}^{(j)}$ directly from the eigenstates of real-space Hamiltonian, and obtain $I_{k}^{(j)}$ via discrete Fourier transformation: 
\begin{equation}
	\phi_{k}^{(j)}(l) = \frac{1}{L} \sum_{p=1}^{L} e^{-i 2 \pi p l / L} \phi_{x}^{(j)}(p).
\end{equation}
Extended eigenstates exhibit $I_{x}^{(j)} \sim L^{-1}$ and $I_{k}^{(j)} \sim \mathcal{O}(1)$, while localized eigenstates have $I_{x}^{(j)} \sim \mathcal{O}(1)$ and $I_{k}^{(j)} \sim L^{-1}$. The averaged IPR over the eigen-spectrum 
\begin{equation}\label{averageIPR}
	\bar{I}_\beta= \frac{1}{L} \sum_{j=1}^{L} I_{\beta}^{(j)}
\end{equation}
can then be used to distinguish different localization phases. In the extended phase, where all eigenstates are extended in the large $L$ limit, $\bar{I}_x \sim L^{-1} \to 0$ while $\bar{I}_k \sim \mathcal{O}(1) \neq 0$. For the localized phase, one has $\bar{I}_x \sim \mathcal{O}(1) \neq 0$ and $\bar{I}_k \sim L^{-1} \to 0$. There is an intermediate phase with coexistence of extended and localized eigenstates separated by mobility edges~\cite{Anderson1958a}, which can be identified by finite values for both $\bar{I}_x$ and $\bar{I}_k$. Figure \ref{fig4}(d) displays $\bar{I}_x$ and $\bar{I}_k$ on the $W$-$m$ plane for $N=L/2=610$. Extended and localized phases are clearly separated by an intermediate region with $\bar{I}_{x,k}\neq0$. The existence of an intermediate phase is further revealed by the mobility edge structure shown in Fig.~\ref{fig4}(e). To identify the boundaries of this phase and others on the $W$-$m$ plane, one can use the quantity $\eta=\log_{10}(\bar{I}_x \times \bar{I}_k)$ \cite{XLi2020a}. For localized or extended phases, $\eta < -\log_{10}L$ because either $\bar{I}_x$ or $\bar{I}_k$ scales as $L^{-1}$. The intermediate phase, however, exhibits larger values of $\eta$ since both $\bar{I}_x$ and $\bar{I}_k$ are finite, as shown in Fig.~\ref{fig4}(f). 

\begin{figure}[tb]
	\centering
	\includegraphics[width=0.45\textwidth]{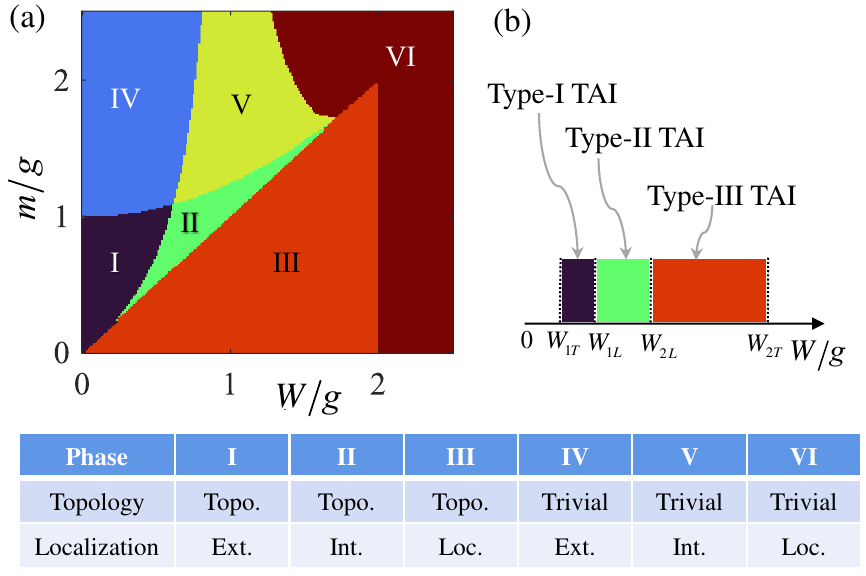}
	\caption{(Color online) (a) Global phase diagram in the $W$-$m$ parameter space, identifying six distinct phase: I: Extended topological phase, II: Topological intermediate phase, III: Topological localized phase, IV: Trivial extended phase, V: Trivial intermediate phase, VI: Trivial localized phase. 
	(b) Disorder strength dependence of three different types of TAIs for fixed $m/g=1.02$.
	Adapted from Ref. \cite{LZTang2022a}.
	}\label{fig5}
\end{figure}

Combining the topological and localization phase diagrams in Figs.~\ref{fig4}(a) and \ref{fig4}(f), one obtains the global phase diagram shown in Fig.~\ref{fig5}(a). This phase diagram comprises six distinct regimes classified by topological and localization characteristics: Phases I--III are respectively topological phases with extended, intermediate, and localized bulk states, while phases IV--VI are trivial counterparts exhibiting corresponding localization properties. This yields three types of gapped TAIs with different bulk states. For example, at $m=1.02$ [Fig.~\ref{fig5}(b)], quasiperiodic disorders can drive transition from trivial extended phase to the type-I, type-II, and type-III TAIs with extended ($W_{1T}<W<W_{1L}$), intermediate ($W_{1L}<W<W_{2L}$), and localized bulk states ($W_{2L}<W<W_{2T}$), respectively. Such TAIs with different bulk states have recently been studied in optical cavity arrays with incommensurate couplings \cite{JFeng2025a}.

\begin{figure*}[tb]
	\centering
	\includegraphics[width=0.8\textwidth]{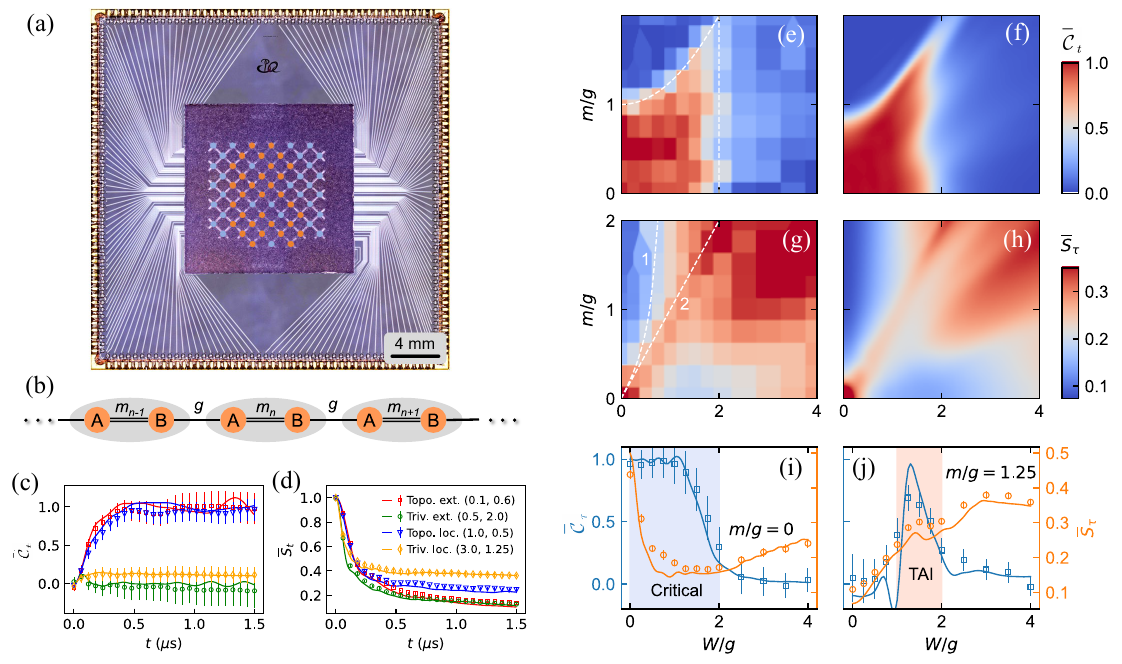}
	\caption{(Color online) (a) Photograph of the programmable superconducting
simulator. (b) Schematic of the gSSH model with quasiperiodic intra-cell hopping. Dynamical evolution of (c) averaged chiral displacement $\overline{\mathcal{C}}_t$ and (d) survival probability $\overline{S}_t$ across distinct topological and localization phases, with parameters $(W/g,m/g)$ specified in the legend. Experimental data (symbols) and numerical simulations (lines) are compared. Topology-localization phase diagram seen from (e,g) experimental and (f,h) numerical results of $\overline{\mathcal{C}}_\tau$ and $\overline{S}_\tau$. Plaquettes in (e,g) represent data points, while white dashed lines indicate theoretical phase boundaries from Fig.~\ref{fig5}. Cross-sections in (i) and (j) display experimental data (symbols) and numerical results (lines), with shaded regions marking pure critical states and TAI regimes. 
	Adapted from Ref. \cite{XGLi2024a}.
	}\label{fig6}
\end{figure*}

\subsection{Other aperiodic systems}
 
The disorder-induced topological phases have been investigated in other aperiodic systems. The TAI with exact mobility edges was found in a 1D SSH chain subjected to slowly varying quasi-periodic modulations \cite{ZPLu2022a}. 
The interplay of quasiperiodic disorders and long-range hopping in the SSH chain can induce multiple TAI phases with different winding numbers \cite{WJZhang2022a}. Similar multiple TAIs with $Z_2$ topological invariants can be induced by Fibonacci modulations in a 1D spin-orbit coupled chain \cite{RJi2025a}. Recent studies on generalized SSH models reveal that the off-diagonal quasiperiodic modulations can induce the reentrant TAI phases \cite{XMWang2025a,ZPLu2025a}, with multiple transitions between distinct TAI states. In the presence of quasiperiodic potentials, a mosaic trimer lattice can also exhibit TAIs and reentrant topological transitions \cite{XTWang2026}. As a spatially short-range correlated disorder, the binary random disorder can lead to the TAIs with reentrant localization transitions in 1D SSH chains \cite{SNLiu2022a,ZWZuo2022a}. In a 1D dimerized Kitaev chain of $p$-wave superconductors, quasiperiodic potential can induce topological phases with critical (multifractal) bulk states \cite{Roy2023a}. In 2D square lattices, it was revealed that a quasiperiodic potential can induce TAIs and topological insulator-to-metal phase transitions through a unique universality class distinct from random systems \cite{YXFu2021a}. In 2D topological models, quasiperiodic potentials can induce Chern insulators \cite{Kuno2019a,Madeira2022a,Goncalves2022a}. 
Unlike uncorrelated disorder, quasiperiodicity can trigger topological transitions through explicit gap closing and reopening, as well as transitions accompanied by the emergence of intermediate metallic and critical phases with unquantized Chern numbers \cite{Goncalves2022a}.

Topological phases in quasicrystals have been explored intensively \cite{Kraus2012a,Kraus2012b,LJLang2012a,Bandres2016a}.
In quasicrystalline systems, disorders can also drive transitions from trivial insulators to topological phases, such as quantum spin Hall states in Penrose and Ammann-Beenker quasicrystals \cite{RChen2019a,TPeng2021a,CBHua2021a}. 
Topological phases can also be realized in amorphous systems with structural disorders \cite{Agarwala2017a,Mitchell2018a,Poeyhoenen2018a,Sahlberg2020a,JYMa2022a}, which means disordered arrangements similar to glass. Recent studies demonstrate that structural disorders can counterintuitively induce topological amorphous phases from trivial crystals \cite{YBYang2019a,JHWang2021a,KLi2021a,CTWang2022a,Corbae2021a}. Amorphous quantum spin Hall states can be induced by structural disorder from quasicrystalline normal insulators \cite{TPeng2024a}. Anderson disorders can also drive amorphous structures into noncrystalline topological insulators \cite{XYCheng2023a} or higher-order topological phases \cite{TPeng2025a}.
In these amorphous systems, the disorder recovers global symmetry under statistical averaging and turns a trivial phase into a topologically nontrivial order \cite{JZhang2022a,RCMa2023a,RCMa2025a}. Such topological states for disordered ensembles are dubbed average symmetry-protected topological phases \cite{RCMa2023a,RCMa2025a}.
Recently, the topological photonic alloy \cite{TTQu2024,TTQu2024a} hosting nontrivial Chern bandgaps and chiral edge states was extended to morphous photonic crystals and quasicrystals \cite{ZYWang2024,BLHuang2025}, where the topology is driven by randomly mixing magnetized and non-magnetized gyromagnetic rods.

\subsection{Experimental realizations} 

\subsubsection{Superconducting quantum simulator}

The gapped TAIs with different bulk states have been recently observed with a programmable superconducting quantum simulator \cite{XGLi2024a}. The quantum simulator has 63 tunable qubits and 105 tunable couplers a 2D square lattice with negligible next-nearest-neighbor coupling, as shown in Fig.~\ref{fig6}(a). A chain of 32 qubits forming a
1D system is used to simulate the generalized SSH model with quasiperiodic intra-cell hopping disorders, described by the Hamiltonian in Eq. (\ref{gSSH}) and shown in Fig.~\ref{fig6}(b). Then the complete topology-localization phase diagram is experimentally mapped out and various topological and trivial phases having extended, critical, and localized bulk states are identified.

The topological and localization properties, characterized by real-space winding number $\nu$ [Eq. (\ref{winding})] and average IPR $\overline{I}_x$ [Eq. (\ref{averageIPR})], are extracted from the time-averaged expectation values of the chiral displacement $\mathcal{C}_t$ and the survival probability $S_t = \frac{1}{t} \int_0^t | \langle l_0 | \psi(t') \rangle |^2  dt'$ using quantum-walk experiments.
Here, the time evolution of the wave function $|\psi(t)\rangle=\exp(-iH_\text{gSSH}t/\hbar)|l_{0}\rangle$, with initial state $|l_{0}\rangle$ localized at site $l_0$. The winding number can be obtained from realization-averaged $\mathcal{C}_t$ \cite{Meier2018a}, while $\overline{I}_x=\lim_{t\to\infty}\overline{S}_{t}$ with $\overline{S}_{t}=\frac{1}{L}\sum_{l_{0}}S_{t}(l_{0})$. 
In the experiment, eight disorder realizations and four
initial single-excitation states for a given target Hamiltonian are constructed to obtain $\overline{\mathcal{C}}_t$ and $\overline{S}_{t}$, with the results for four representative $(W/g, m/g)$ parameter configurations shown in Figs.~\ref{fig6}(c) and \ref{fig6}(d). The quantum-walk dynamics confirm that edge excitations maintain spatial confinement regardless of bulk state localization in topological phase, while only localized states preserve edge confinement and extended states exhibit complete delocalization across all sites in trivial phase. In addition, $\overline{\mathcal{C}}_{t}$ asymptotically converges to quantized values reflecting the topological invariant $\nu=1$ for topological phase and $0$ for trivial phase. The decay rate of $\overline{S}_{t}$ in the extended phase substantially exceeds that in the localized phase, and $\overline{S}_{t}$ saturates to distinct values for discriminating localized and extended states. The quantities $\overline{\mathcal{C}}_\tau$ and $\overline{S}_\tau$, measured at the maximal evolution time $\tau$, serve as experimental proxies to map the topology-localization phase diagram in the $W$-$m$ parameter plane, as displayed in Figs.~\ref{fig6}(e) and \ref{fig6}(g). Numerical simulations in Figs.~\ref{fig6}(f) and \ref{fig6}(h) show quantitative consistency between experiment and theory \cite{LZTang2022a}. As shown in Figs.~\ref{fig6}(i) and \ref{fig6}(j) for $m/g=0$ and $1.25$, the pure critical states and TAI regimes are experimentally indicated, respectively.

\subsubsection{Photonic lattice}

Recently, the gapped TAI phase has also been realized in a quasiperiodic photonic lattice \cite{Sinha2025a}. The gSSH model with quasiperiodic intra-cell hopping disorder, described by Eq. (\ref{gSSH}), is implemented in a femtosecond laser-written lattice of 80 sites with tunable inter-waveguide coupling based on the wavelength-tuning technique. By measuring the mean chiral displacement from the transport of light in the bulk of the lattices, the topological Anderson transition caused by increasing quasiperiodic disorder strength is dynamically probed \cite{Sinha2025a}: The system exhibits a reentrant transition from a trivial phase to the TAI and back to a trivial phase, accompanied by the closing and reopening of the band gap.

Notably, a gapless TAI phase was theoretically proposed in a SSH waveguide lattice with quasiperiodic disorders in inter-cell couplings \cite{Longhi2020a}. In this TAI phase, the zero-energy edge modes reside within the gapless localized bulk states, which is in contrast to the gapped TAIs with various localization properties in Ref. \cite{LZTang2022a}. Such a quasiperiodic SSH lattice with inter-cell disorders has been engineered in a laser-written waveguide array \cite{WCheng2022} and a nanophotonic circuit \cite{JGao2022}, respectively. The spectral analysis based on the continuous measurement of the light dynamics near edge lattices has been used to detect the disorder-induced zero-energy edge modes \cite{WCheng2022,JGao2022}, as suggested in Ref. \cite{Longhi2020a}.

\section{Non-Hermitian TAIs}
In recent years, growing efforts are devoted to exploring topological phases in non-Hermitian systems, including new topological invariants \cite{Ghatak2019a}, the non-Hermitian skin effect and modified bulk-boundary correspondence \cite{SYYao2018a,Lee2016a}. In addition, non-Hermitian systems are found to exhibit distinctive localization behaviors in the presence of disorders, which open new avenues for studying disordered topological phases beyond the Hermitian framework. In this section, we will focus on the non-Hermitian TAIs introduced in Refs. \cite{DWZhang2020a,XWLuo2023a,LZTang2020a} and the disorder-induced point-gap topology introduced in Ref. \cite{Claes2021a}. We also review relevant theoretical and experimental studies of these topological phases in (artificial) non-Hermitian disordered systems.

\subsection{One dimension}

\begin{figure}[tb]
	\centering
	\includegraphics[width=0.46\textwidth]{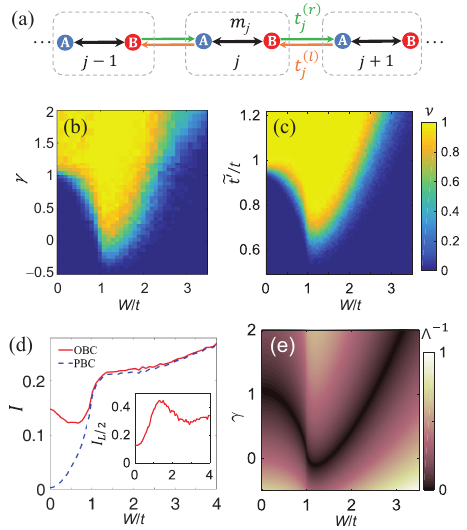}
	\caption{(Color online) (a) Sketch of the non-Hermitian disordered SSH model in Eq. (\ref{nHSSH1}). The dotted box is the unit cell, and $m_j$ and $t_j^{(l,r)}$ are Hermitian intra-cell and nonreciprocal inter-cell hoppings. (b, c) Winding number $\nu$ versus disorder strength $W$ and non-Hermiticity $\gamma$ (b) or renormalized hopping $\tilde{t}^{\prime}$ (c). (d) $\bar{I}$ and $I_{L/2}$ (inset for $L=400$) as a function $W$ and fixed $\gamma=0.6$. (e) $\Lambda^{-1}$ as a function of $\gamma$ and $W$. Other parameters are $t^{\prime} = 0.7t$, $l = 0.2L$, $W_1 = W$, $W_2 = 0$, $N_s = 50$. Adapted from Ref. \cite{DWZhang2020a}.
	}\label{fig7}
\end{figure}

Let us consider a nonreciprocal disordered SSH model, as shown in Fig.~\ref{fig7}(a). The tight-binding Hamiltonian reads \cite{DWZhang2020a}
\begin{equation}\label{nHSSH1}
	H=\sum_j \left[ \left(m_ja_j^\dagger b_j + \mathrm{H.c.}\right) + t_j^{(r)}a_{j+1}^\dagger b_j + t_j^{(l)}b_j^\dagger a_{j+1}  \right] ,
\end{equation}
where $m_{j} = t + W_{1} \omega_{j}$, $t_{j}^{(l)} = t' + W_{2} \omega_{j}'$, and $t_{j}^{(r)} = t_{j}^{(l)} + t^{\prime}\gamma$ with $\gamma$ as the non-Hermiticity strength in this nonreciprocal term. Here $a_j^\dagger$ ($b_j^\dagger$) and $a_j$ ($b_j$) create and annihilate particles on sublattice A (B) in the unit cell $j$. The intra-cell and inter-cell hopping strengths are denoted by $t$ and $t^\prime$, respectively. Independent random variables $\omega_j$ and $\omega_j^{\prime}$ are uniformly distributed in $[-1,1]$, with $W_1$ and $W_2$ representing their random disorder strengths. We set $t\equiv 1$ as the energy unit, and consider the case of $W_2 = 0$ and $W_1=W$ here (see Ref. \cite{DWZhang2020a} for the cases with finite $W_2$). This system preserves chiral symmetry $\Gamma H \Gamma^{-1}=-H$ with the chiral operator $\Gamma=\sigma_z\otimes\mathbb{I}$, where $\sigma_z$ and $\mathbb{I}$ denote Pauli matrix and identity matrix, respectively. One can extend the real-space winding number in Eq. (\ref{winding}) to the non-Hermitian disordered SSH model. For a disorder configuration labeled $s$, the Hamiltonian (\ref{nHSSH1}) under OBCs can be diagonalized into two chiral-symmetric sectors: $H_s|nR_\pm\rangle_s = \pm E_{n,s}|nR_\pm\rangle_s$, where $|nR_-\rangle_s = \Gamma|nR_+\rangle_s$. In the biorthonormal basis for non-Hermitian systems, the corresponding left eigenstates $|nL_\pm\rangle$, orthonormal to the right eigenstates, are constructed from the columns of $(T_s^{-1})^\dagger$ via the diagonalization $H_s = T_s \Lambda_s T_s^{-1}$. The open-boundary flat-band Hamiltonian $Q_s$ under OBCs is defined homotopically as $Q_s = \sum_{n=1}^L \left( |nR_+\rangle_{ss}\langle nL_+| - |nR_-\rangle_{ss}\langle nL_-| \right)$. The real-space winding number for a single disorder realization $\nu_s$ is then given by $\nu_s = \frac{1}{2L^{\prime}} \mathrm{Tr}^{\prime}(\Gamma Q_s [Q_s, X])$, with $X$ the coordinate operator, $\mathrm{Tr}^{\prime}$ the trace over the central segment of length $L^\prime$. The disorder-averaged winding number reads $\nu = \frac{1}{N_s} \sum_{s=1}^{N_s} \nu_s$, where a modest number $N_s$ of configurations are sufficient in practice.

We investigate the interplay of non-Hermiticity and disorder in an initially trivial SSH chain with $t > t'$. The disorder-averaged winding number $\nu$ is numerically evaluated in $\gamma-W$ plane, as shown in Fig.~\ref{fig7}(b). In the clean Hermitian limit ($\gamma = W = 0$), the system remains trivial with $\nu \simeq 0$. Under moderate non-Hermiticity and disorder, a regime with $\nu \approx 1$ emerges, which approaches unity with increasing the system size. In this topological regime, two zero-energy edge modes are exhibited without bulk gaps under OBCs. This indicates that a topological insulating phase can be induced from an initially trivial Hermitian clean system by combining moderate non-Hermiticity and disorder, which is thus termed as a non-Hermitian TAI (NHTAI) \cite{DWZhang2020a}. To elucidate the connection between NHTAIs and their Hermitian counterparts, one can employ a similarity transformation that maps the non-Hermitian open SSH chain to a Hermitian one with a renormalized inter-cell hopping amplitude $\tilde{t}^{\prime} = t^{\prime} \sqrt{1 + \gamma}$. Using the right eigenstates of this effective Hermitian Hamiltonian, one can compute the real-space winding number in Fig.~\ref{fig7}(c). One can see that a TAI phase emerges in the region $0.7 < \tilde{t}^{\prime}/t < 0.97$, which closely aligns with the NHTAI region for $0 < \gamma < 0.95$ in the non-Hermitian model.

Although the NHTAI phase can be topologically connected to the TAI in this case, the NHTAI has unique properties without Hermitian counterparts and even exists without the similarity transformation when the energy spectrum
is generally complex \cite{DWZhang2020a}. For instance, due to the non-Hermitian skin effect \cite{SYYao2018a,Kunst2018a}, the NHTAI has unconventional bulk-edge correspondence and non-monotonous localization behavior under OBCs. The spectrum-averaged IPR $\bar{I}$ and the IPR for $L/2$-th eigenstate $I_{L/2}$ with $L=400$ as a function of disorder strength $W$ are shown in Fig.~\ref{fig7}(d). 
Under OBCs, the global localization index $\bar{I}$ varies non-monotonously for $W \lesssim t$. This behavior originates from the interplay between Anderson localization and the NHSE. In contrast, bulk states under PBCs become monotonously localized.
The values of $I_{L/2}$ under OBCs are larger in the NHTAI phase regime implies that the zero-energy edge modes are more localized. Under strong disorders, the skin effect is destroyed and the values of $\bar{I}$ for OBCs and PBCs are almost the same. The topological critical points can be identified from the divergence of the localization length of zero-energy states $\Lambda$ (i.e., $\Lambda^{-1}\rightarrow0$), as shown in Fig.~\ref{fig7}(e). This indicates that the topological transition is accompanied by an Anderson localization-delocalization transition in the non-Hermitian system.

\begin{figure}[tb]
	\centering
	\includegraphics[width=0.46\textwidth]{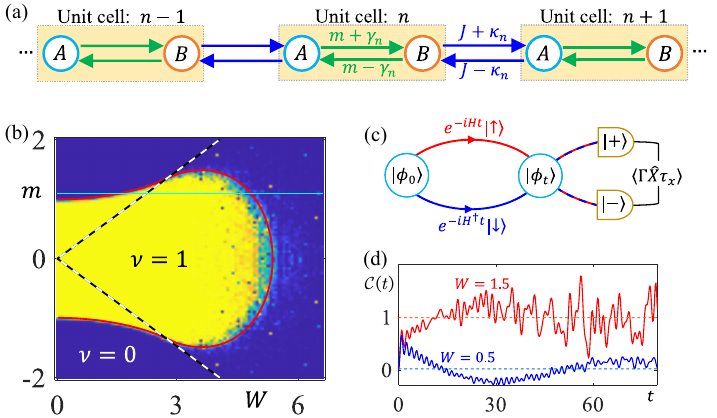}
	\caption{(Color online) (a) Sketch of the generalized SSH model in Eq. (\ref{nHSSH2}) with purely non-Hermitian disorders $\gamma_n$ and $\kappa_n$. (b) Phase diagram in the $m-W$ plane with $\kappa_{b}=0.1$, obtained from the winding number $\nu$. Red solid line is the analytic phase boundary, and striped line denotes the PT-symmetry breaking curve. (c) Ramsey interferometer for measuring the biorthogonal chiral displacement $\mathcal{C}(t)$. (d) Dynamics of $\mathcal{C}(t)$ for topological ($W=1.5$) and trivial ($W=0.5$) phases.  Adapted from Ref. \cite{XWLuo2023a}.
	}\label{fig8}
\end{figure}

The 1D NHTAI phase has also been proposed in the generalized SSH model with purely non-Hermitian disorders \cite{XWLuo2023a}. As shown in Fig.~\ref{fig8}(a), the model is described by the tight-binding Hamiltonian \cite{XWLuo2023a}:
\begin{equation}\label{nHSSH2}
	H=\sum_n \left(c_n^\dagger M_nc_n+J_n^+c_n^\dagger\sigma^+c_{n+1}+J_n^-c_{n+1}^\dagger\sigma^-c_n \right).
\end{equation}
Here, $c_n^{\dagger}=(c_{n,A}^{\dagger},c_{n,B}^{\dagger})$ is the particle creation operator of sites $A$ and $B$ in the unit cell $n$, $M_n=m\sigma_x-i\gamma_n\sigma_y$, and $\sigma^\pm=\frac12(\sigma_x\pm i\sigma_y)$, $J_n^\pm=J\pm\kappa_n$, with $m$ and $J$ being the uniform Hermitian parts of the intra- and inter-cell tunnelings, letting $J\equiv1$ as the energy unit. The anti-Hermitian parts $\gamma_{n}$ and $\kappa_{n}$ give the purely non-Hermitian disorders, which are independently and uniformly distributed in the range $[-\frac{W_{\gamma}}{2},\frac{W_{\gamma}}{2}]$ and $[\kappa_{\mathrm{b}}-\frac{W_{\kappa}}{2},\kappa_{\mathrm{b}}+\frac{W_{\kappa}}{2}]$ with a bias $\kappa_\mathrm{b}$, respectively. Here $W_{\gamma}$ and $W_{\kappa}$ denote the random disorder strengths. Apart from the chiral symmetry, this model preserves a hidden PT symmetry that can be broken by strong non-Hermitian disorders. Under stronger intra-cell disorders, such as $W_{\gamma}=8W_{k}=W$ in Fig.~\ref{fig8}(b), the topological phase diagram in the $W$-$m$ plane is obtained by the real-space winding number $\nu$. The non-Hermitian disorders enlarge the topological region with $\nu=1$, and can drive the transition from a trivial phase in the clean limit to the NHTAI as $W$ increases, as shown along the horizontal line in Fig.~\ref{fig8}(b) with $m=1.1$. In large $W$ limit, the intra-cell couplings dominate and the system becomes trivial again. It has been further proposed that the winding number for the 1D NHTAI can be extracted by measuring the biorthogonal chiral displacement $\mathcal{C}(t)$ \cite{XWLuo2023a}, based on a proper Ramsey interferometer as shown in Fig.~\ref{fig8}(c). An initial state $|\phi_0\rangle$ with an additional pseudospin degree of freedom is prepared and then evolves with two components $|\uparrow\rangle$ and $|\downarrow\rangle$ governed by Hamiltonians $H$ and $H^\dagger$, respectively. After an evolution interval $t$, the biorthogonal chiral displacement in basis $|\pm\rangle$ can be measured. As shown in Fig.~\ref{fig8}(d), the disorder- and time-averaging $\mathcal{C}(t)$ converge to 0 and 1 for trivial and topological phases, respectively.

The topological transition driven by loss-and-gain disorders has also been studied in a dimerized Kitaev superconductor chain \cite{CHZhang2021a}. The topological superconducting phase with Majorana zero-energy edge modes can be induced by the non-Hermitian disorder, which provides a non-Hermitian extension of the topological Anderson superconductor in Hermitian systems \cite{Borchmann2016a,WQin2016a,Habibi2018,Lieu2018a,CHua2019a}. In addition, the topological Anderson transitions in a non-Hermitian generalized SSH model with nonreciprocal hopping and complex quasiperiodic hopping modulation was studied in Ref. \cite{XQSun2023a}, where two delocalization transitions without accompanying the topological transition were found.

\subsection{Two dimensions}

\begin{figure}[tb]
	\centering
	\includegraphics[width=0.45\textwidth]{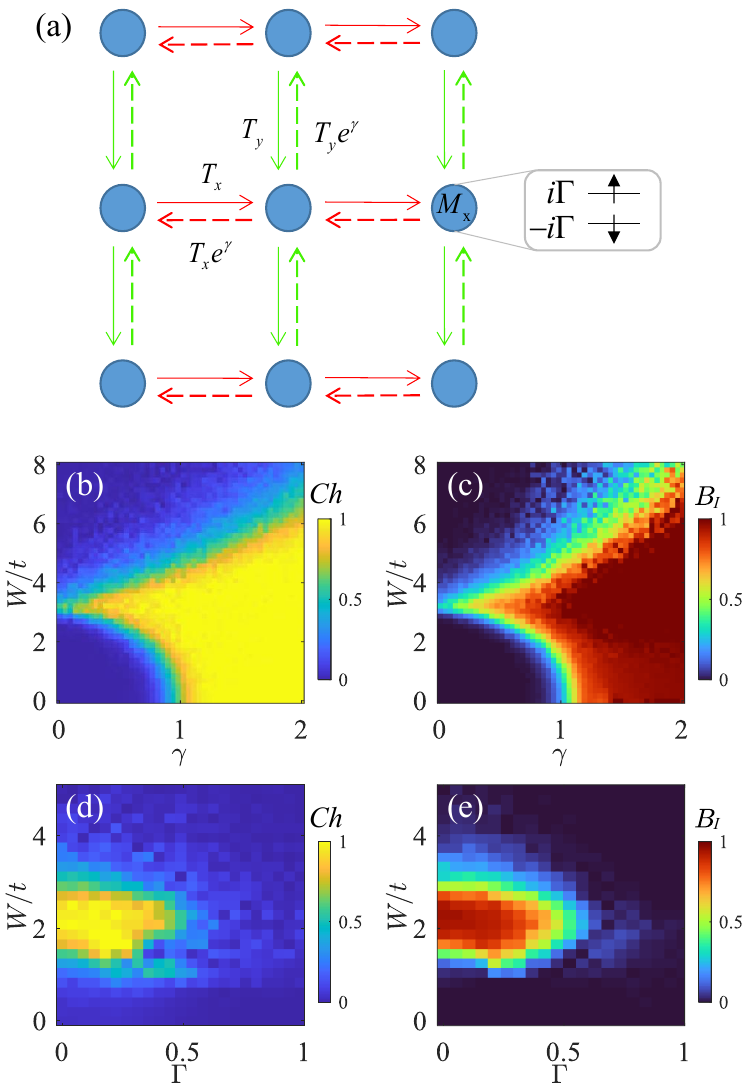}
	\caption{(Color online) (a) Schematic of the non-Hermitian disordered Chern-insulator model with the site index $\mathbf{x}=(x,y)$. Here $T_{x,y}$ and $T_{x,y}e^\gamma$ denote the nonreciprocal intercell hopping matrices on the (pseudo)spin basis $\{\uparrow,\downarrow\}$ along the $x,y$ axis, and $M_\mathbf{x}$ denotes a disordered on-site Zeeman potential with a gain-and-loss term. (b) The Chern number ${Ch}$ and (c) Bott index $B_I$ as a function of the nonreciprocal parameter $\gamma$ and the disorder strength $W$ for $\Gamma=0$ and $m=3$. (d) ${Ch}$ and (e) $B_I$ as a function of the non-Hermitian gain-and-loss parameter $\Gamma$ and the disorder strength $W$ for $\gamma=0$ and $m=2.2$. Other parameter throughout is $L=16$.
	Adapted from Ref. \cite{LZTang2020a}.
	}\label{fig9}
\end{figure}

The model of Chern insulators on a square lattice can be generalized to the  non-Hermitian regime, as shown in Fig.~(\ref{fig9}). The non-Hermiticity can be introduced through the nonreciprocal intercell hopping and an on-site gain–loss term incorporated into the Zeeman potential. The tight-binding Hamiltonian reads \cite{LZTang2020a}
\begin{equation}\label{LZTang2020a_H}
	H=\sum_\mathbf{x}\sum_{j=x,y}\left(c_\mathbf{x}^\dagger T_jc_{\mathbf{x}+\mathbf{e}_j}+c_{\mathbf{x}+\mathbf{e}_j}^\dagger T_j^\dagger e^\gamma c_\mathbf{x}\right)+\sum_\mathbf{x}c_\mathbf{x}^\dagger M_\mathbf{x}c_\mathbf{x},
\end{equation}
where $c_{\mathbf{x}}^{\dagger}=(c_{\mathbf{x},\uparrow}^{\dagger},c_{\mathbf{x},\downarrow}^{\dagger})$ represents a two-component creation operator for a particle at lattice site $\mathbf{x}=(x,y)$ with (pseudo)spin states $\{\uparrow,\downarrow\}$. The hopping matrices are given by $T_j=-\frac12t_j \sigma_z-\frac i2v_j \sigma_j$, and the on-site potential takes the form
$M_{\mathbf{x}}=(m_{\mathbf{x}}+i\Gamma)\sigma_{z}$. Here, $\gamma$ and $\Gamma$ denote the strengths of non-Hermiticity arising from
nonreciprocal hopping and the gain and loss, respectively. For convenience, we adopt the parameter values $t_x=t_y=t=1$ and $v_x=v_y=v=1$, with $t=1$ as the unit of energy. Diagonal disorder is incorporated into the Zeeman term via $m_{\mathbf{x}}=m+W\omega_{\mathbf{x}}$, where $m$ is a uniform Zeeman strength and $\omega_\mathrm{x}$ are independent random variables uniformly distributed in the range $[-1,1]$, with $W$ as the disorder strength. In the Hermitian and clean limit, the Hamiltonian Eq. (\ref{LZTang2020a_H}) reduces to a square-lattice realization of a 2D Chern insulator, representing a variant of the Haldane model \cite{Haldane1988a}. In this case, the system exhibits a nonzero Chern number for $|m| < 2$, accompanied by chiral edge modes propagating along the boundary under OBCs. When $|m| > 2$, both energy bands become topologically trivial with zero Chern number.

One can extend the real-space Chern number ${Ch}$ and Bott index $B_I$ with biorthonormal formula to characterize the topology of this non-Hermitian 2D disordered system \cite{FSong2019a,LZTang2020a}. Here we focus on the effect of combining disorder and non-Hermiticity on a trivial phase at the clean and Hermitian limit. For the nonreciprocal hopping case with fixed $\Gamma=0$ and $m=3$, the topological phase diagrams in the $\gamma$-$W$ plane are obtained by computing the disorder-averaged ${Ch}$ and $B_I$, as shown in Figs.~\ref{fig9}(b) and \ref{fig9}(b), respectively. One can see the nonreciprocal hopping $\gamma$ can enlarge the topological region, which includes the phase transition from the clean trivial phase to the topological phase when $W\lesssim6$. Thus, the NHTAI phase can be induced by the combination of modest disorder and non-Hermiticity in this 2D system. For the gain-and-loss case, the numerical results of ${Ch}$ and $B_I$ as a function of $\Gamma$ and $W$ with fixed $\gamma=0$ and $m=2.2$ are shown in Figs.~\ref{fig9}(d) and \ref{fig9}(e), respectively. One can find that the gain-and-loss $\Gamma$  reduces the topological regions in the phase diagrams. In the Hermitian limit, the TAI exists under modest disorder strength $W$. As $\Gamma$ increases, the TAI persists when $\Gamma\lesssim0.5$ and becomes trivial under larger $\Gamma$. Since the nonreciprocal hopping enlarges the topological region, one can expect the 2D NHTAI phase to persist in this non-Hermitian lattice with both finite $\gamma$ and $\Gamma$. The localization properties and related expansion dynamics under the interplay of disorder and non-Hermiticities in this system have also been explored \cite{LZTang2020a}.

The 2D NHTAI phase has also been found in the non-Hermitian disordered square lattice with imaginary Zeeman fields and on-site random disorders \cite{HFliu2020a}. The topological phase transitions, which are characterized by the non-Hermitian extension of quadrupole moments \cite{HWu2021a}, can be induced by imaginary (gain and) loss disorders in 2D non-Hermitian lattices \cite{HFLiu2021a,QYMo2022a}. Moreover, it was revealed that adding sufficient imaginary disorder into a trivial insulator can lead to a higher-order TAI hosting quantized quadrupole moment and corner states \cite{CALi2020a,YBYang2021a,YBYang2024a,Wenbin2025a}, termed as non-Hermitian higher-order TAIs \cite{HFLiu2021a,QYMo2022a}. Recently, the disorder-induced exceptional points accompanied by topological phase transitions were shown in a 2D non-Hermitian lattice with non-reciprocal and random hoppings \cite{XYCheng2025a}. The random disorder can drive this system from a trivial insulator to a NHTAI at the exceptional points.

\subsection{Experimental realizations}

\begin{figure}[tb]
	\centering
	\includegraphics[width=0.45\textwidth]{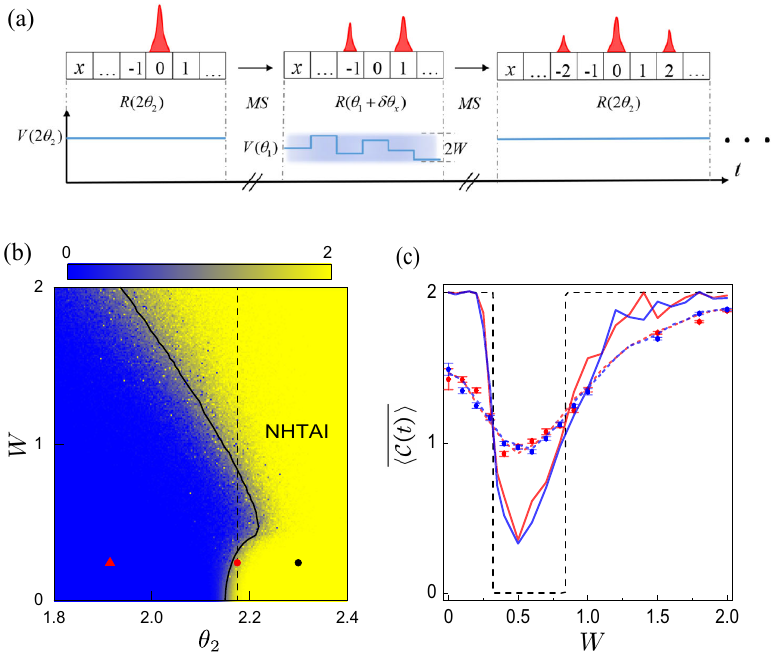}
	\caption{(Color online)  (a) Illustration of the operation
sequence of the time-multiplexed quantum walk of single photons for observing NHTAI. (b) Theoretical phase diagram in terms of the color contour of the numerically evaluated topological marker, with the yellow (blue) region for the topologically non-trivial (trivial) phase. The labeled NHTAI state corresponds to the yellow region with $W>0$. (c) Measured time and disorder-averaged biorthogonal chiral displacement for 9-step quantum walks (vertical dashed line in (b)). Experimental (numerical) data were represented by blue and red dots (solid lines) for $\gamma=0$ and $\gamma=0.1$, respectively. Blue and red solid lines are numerically evaluated chiral displacement for 400-step quantum walks.
	Adapted from Ref. \cite{QLin2022a}.
	}\label{fig10}
\end{figure}

\begin{figure*}[t]
	\centering
	\includegraphics[width=0.95\textwidth]{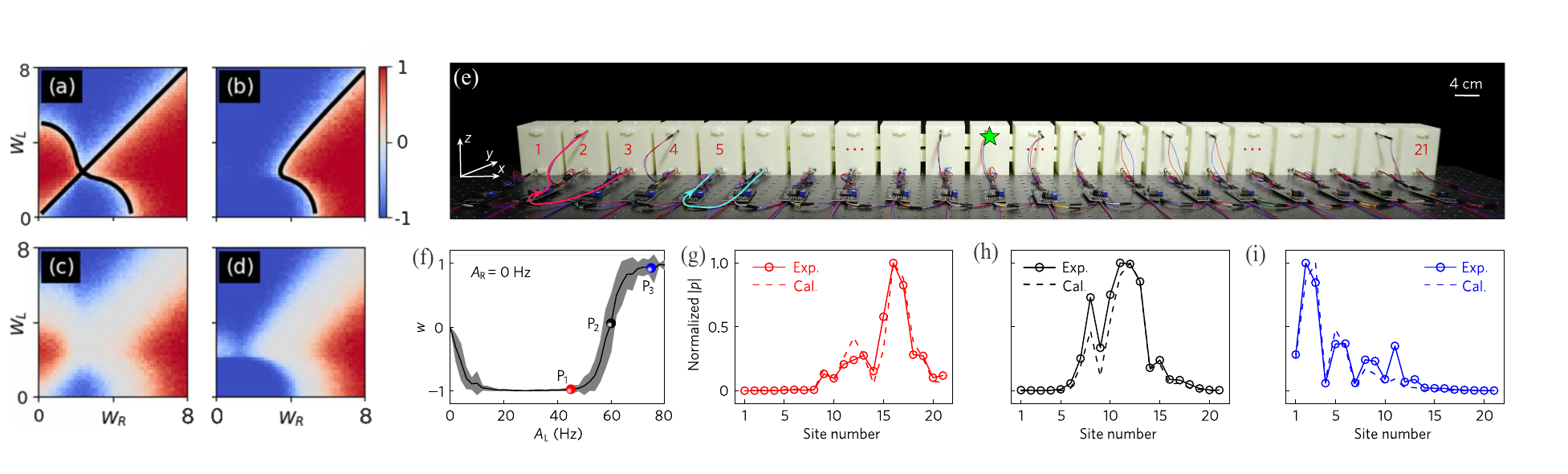}
	\caption{(Color online) Point-gap winding number $w(0)$ as a function of $(W_R,W_L)$ for $(J_L,J_R,W)=$ (1,1,0) (a), $(J_L,J_R,W)=(1,0.5,0)$ (b), $(J_L,J_R,W)=(1,1,1)$ (c), and $(J_L,J_R,W)= (1,0.5,1)$ (d). The black lines denote the points where the localization length of zero-energy states diverges. (a-d) Adapted from Ref. \cite{Claes2021a}. (e) Photo of a 1D disordered acoustic crystal with positive and negative left-directional nearest-neighbor couplings, provided by externally connected sets (red and cyan curves). The green star denotes the sound source. (f) Calculated $w(0)$ of the disordered acoustic crystal with $A_{\mathrm{R}}=0$ Hz. The points $P_{1}$, $P_{2}$, and $P_{3}$ correspond to $A_{\mathrm{L}}=45$ Hz, $60$ Hz, and $75$ Hz, respectively. (g-i) Experimentally measured and calculated sound pressure ($|p|$) distributions at points $P_{1}$ (g), $P_{2}$ (h), and $P_{3}$ (i). (e-i) Adapted from Ref. \cite{BBWang2025a}.
	}\label{fig11}
\end{figure*}

The NHTAIs have recently been observed in several artificial non-Hermitian systems. The first experiment to simulate the 1D NHTAI is using disordered single-photon quantum-walk dynamics \cite{QLin2022a}. Photons are sent through an interferometric network consisting of optical elements. As illustrated in Fig.~\ref{fig10}(a), the Floquet operator with a time-multiplexed sequence acts on the walker state of photons. The walker position can be mapped to the time domain and two polarizations of photons act as pseudospins. 
By adding proper polarization-dependent loss operations and position-dependent 
rotation modulations \cite{QLin2022a}, the tunable non-Hermiticity and random hopping disorder are respectively realized in this quantum dynamics system. The system (Floquet operator) has the chiral symmetry with the topological phase diagram shown in Fig.~\ref{fig10}(b), evaluated by the disorder-averaged biorthogonal winding number. The NHTAI state corresponds to the yellow region with the finite disorder in the phase diagram. By measuring the time and disorder-averaged biorthogonal chiral displacement \cite{QLin2022a,XWLuo2023a}, the topological nature of the NHTAI state was probed, as shown in Fig.~\ref{fig10}(c). The non-monotonous localization behaviour \cite{DWZhang2020a}, which is due to the competition between Anderson localization and skin effect, has also been experimentally revealed through dynamic observables.

The NHTAIs have also been observed in classical wave systems, such as acoustic lattices \cite{ZMGu2023a,QYMo2022a} and heat transport system \cite{HGao2024a}. In Ref. \cite{ZMGu2023a}, the NHTAI and related topological phase transitions solely driven by disordered loss modulation in a 1D acoustic lattice were experimentally revealed using both bulk and edge spectra. In Ref. \cite{QYMo2022a}, a reconfigurable 2D acoustic lattice with random loss configuration was fabricated to demonstrate the non-Hermitian higher-order TAI with corner states induced solely by imaginary disorders. The non-Hermitian higher-order TAI has also been observed in a 2D disordered convective lattices with thermal dissipation \cite{HGao2024a}, which is revealed by the emergence of corner states with energy localizations as the disorder strength increases.

\subsection{Disorder-induced point-gap topology}

The non-Hermitian skin effect is an interesting phenomenon unique to non-Hermitian systems \cite{SYYao2018a}, which manifests as a collapse of extended bulk modes into a macroscopic number of boundary-localized states under OBCs. This phenomenon is related to the point-gap topology of complex energy spectra, which is characterized by a non-Hermitian winding number defined in the complex energy plane \cite{Okuma2020a}. 
{ For a 1D non-Hermitian Hamiltonian $H$, the point-gap is defined when the complex energies doesn't cross a reference point $E_0\in \mathbb{C}$ in the complex plane, enabling the definition of a non-Hermitian winding number $w(E_0)$ \cite{Okuma2020a}
\begin{equation}
	w(E_0) = \int_0^{2\pi} \dfrac{d\theta}{2\pi i} \dfrac{d}{d\theta} \log \det (H(\theta)-E_0).
\end{equation}
Here $\theta \in [0, 2\pi]$ denotes a phase twist (magnetical flux) applied to the PBC rings.
The point gap is distinct from a line gap, where the spectrum is partitioned into two disconnected regions by a straight line. It is important to distinguish these from the conventional bulk gap in Hermitian systems: while the latter is essentially a line gap restricted to the real energy axis, a non-Hermitian line gap can be generalized to the imaginary axis or any arbitrary line in the complex plane. In contrast, a point gap does not require such a spectral separation but instead focuses on the global winding of the complex energy bands around a base point. }
{ Building on these foundations, various disorders have been shown to trigger point-gap transitions and spectral deformations \cite{Longhi2021, ZHWang2021, JRLi2024, WWang2025}. Such systems exhibit a multifaceted competition between the NHSE and Anderson localization\cite{JRLi2026, JRLi2026a}.}
Notably, a skin effect is developed by adding disorder to a clean system, termed the non-Hermitian Anderson skin effect (NHASE) \cite{Claes2021a}. The NHASE can be regarded as a non-Hermitian counterpart of the TAI, wherein the skin effect emerges solely due to disorder in a system that originally exhibits no skin effect in the clean limit. These findings reveal a mechanism of disorder-induced localization with non-Hermitian topology beyond the conventional framework of Anderson localization in Hermitian systems.

To illustrate the NHASE, the authors in Ref. \cite{Claes2021a} consider the generalized Hatano-Nelson model \cite{Hatano1996a,Hatano1997a}:
\begin{equation} \label{GHN} {H}=\sum_iJ_R^i{c}_{i+1}^\dagger{c}_i+J_L^i{c}_i^\dagger{c}_{i+1}+h^i{c}_i^\dagger{c}_i.
\end{equation}
In this model, $J_L^i=J_L+W_L\omega_L^i$ and $J_R^i=J_R+W_R\omega_R^i$ denote the nonreciprocal disordered hoppings in the left and right directions, $h^i=W\omega^i$ is an on-site potential. Here $\omega_{(L/R)}^i\in[-0.5,0.5]$ are uniformly distributed random variables, and $W$ controls the random disorder strength. To study the point-gap topology in this disordered non-Hermitian system, a real-space formula of the complex-energy winding number $w(E_0)$ is developed \cite{Claes2021a}. Figures \ref{fig11}(a) and \ref{fig11}(b) show the numerically computed $w(0)$ as a function of the disorder parameters $(W_{L},W_{R})$, for fixed parameters (a) $(J_L,J_R,W)=(1,1,0)$ and (b) $(J_L,J_R,W)=$ (1,0.5,0). One can find that $w(0)$ is quantized and only changes when the zero-energy states have diverging localization length (black lines). In Figs.~\ref{fig11}(c) and \ref{fig11}(d), the results of $w(0)$ for systems with (c) $\left(J_{L},J_{R},W\right)= (1,1,1)$ and (d) $(J_L,J_R,W)=(1,0.5,1)$ are shown. The skin effect occurs at the left (right) edge when the point-gap winding number $w<0$ ($w>0$), and no skin effect occurs when $w=0$. This connection allows one to reveal the NHASE, in which the system develops a skin effect at a critical value of disorder. For instance, the system in the clean case ($W_{L,R}=0$) has no NH skin effect with $w=0$ in Fig.~\ref{fig11}(c), but transitions to $w=\pm1$ at nonvanishing critical values of $W_L$ or $W_R$.


The NHASE with disorder-induced point-gap topology has been comprehensively studied for 1D non-Hermitian systems with different symmetry classes \cite{Sarkar2022a}. The generalized Brillouin zone theory was developed to reconstruct the bulk-boundary correspondence and precisely describe various skin effects in disordered non-Hermitian systems \cite{ZQZhang2023a}, including the disorder-induced NHASE. 
A second-order NHASE with skin modes localized at the corners under OBCs was shown by adding disorders to a trivial non-Hermitian 2D system \cite{Kim2021a}. Recently, using an electrical-circuit analog, it was theoretically and experimentally demonstrated that coupled non-Hermitian and Hermitian chains antisymmetrically correlated disorders exhibit a topological Anderson transition with reemergent NHASE \cite{WWJin2025a}.

The NHASE with disorder-induced point-gap topology has recently been observed in a non-Hermitian acoustic crystal \cite{BBWang2025a}. As schematically
shown in Fig.~\ref{fig11}(e), the resonators and small tubes serve as sites and couplings of the lattice, adjacent resonators externally connected to the amplifiers are used to realize disordered unidirectional couplings.
In this setting, one can realize a simplified version of the 1D tight-binding model in Eq. (\ref{GHN}) with $h^i=0$ (no on-site potential) and redefined nonreciprocal disordered hoppings $J_{R(L)}^i=t+A_{R(L)}r_{R(L)}^i$, where random numbers $r_{R(L)}^i\in[-0.5,0.5]$ and $A_{R(L)}$ denote the disorder strengths. Figure \ref{fig11}(f) plots the calculated point-gap winding number as a function of $A_{L}$ with $A_{R}=0$ for the acoustic crystal. Three phases exist with the winding numbers -1, 0, 1, which correspond to the right-skin, no-skin, and left-skin states, respectively. By measuring the pressure amplitude distributions in the acoustic crystal, this disorder-induced topological transition process of the skin effect was observed from the sound field, as shown in Figs.~\ref{fig11}(g-i). The left or right boundary where localization occurs depends on the strength of the disorder. As the disorder strength increases, the direction of boundary localization can be reversed.

\section{Disorder-induced topology in dynamical systems}

In this section, we review the disorder-induced topological effects in dynamical systems. We introduce the topological Anderson-Thouless pump and its experimental realization, and other disorder-induced topological pumps in periodically driven systems.

\subsection{Topological Anderson-Thouless pump}

\begin{figure*}[tb]
	\centering
	\includegraphics[width=0.85\textwidth]{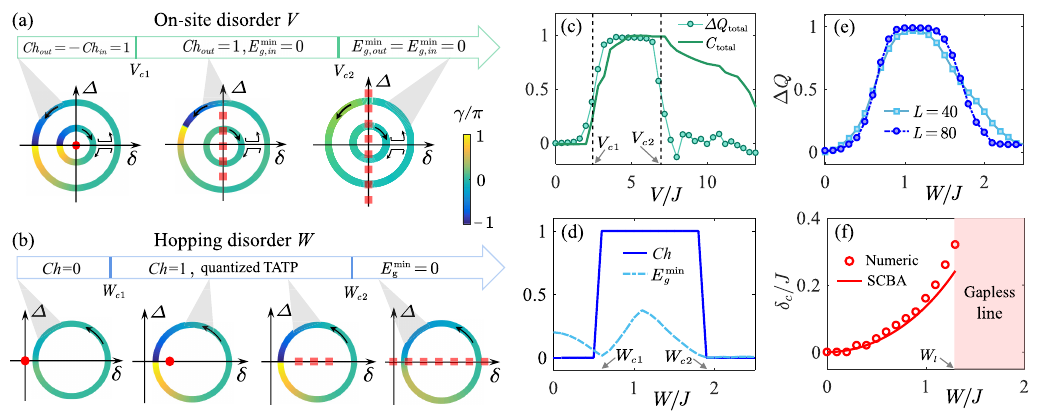}
	\caption{(Color online) Topological pumps driven by (a) on-site disorder $V$ and (b) hopping disorder $W$. Pump loops in the $\delta$-$\Delta$ plane are shown, with colors and arrows indicating the Berry phase $\gamma$ and pumping direction; red markers and dashed curves denote gapless points and lines, respectively. Critical disorder strengths are $\{V_{c1},V_{c2}\}\approx\{2.5,6.8\}$ and $\{W_{c1},W_{c2}\}\approx\{0.6,1.9\}$.
	(c) Pumped charge $\Delta Q_\mathrm{total}$ and Total Chern number ${Ch}_\mathrm{total}$ versus $V$ for $L=80$, $T=400$. Dashed lines indicate $V_{c1}$ and $V_{c2}$, where each loop crosses a gapless region. (d) Chern number ${Ch}$ and minimum gap $E_g^\mathrm{min}$ versus $W$. (e) $\Delta Q$ versus for two system sizes and periods. (f) The disorder-induced shift of the gapless point $\delta_c$ for $W<W_l\approx1.3$.
	Other parameters: $J=1,\beta=0,U=0;$ loop parameters are $\{R_\Delta,R_\delta,\delta_s\}=\{5,5,0\}$,
	$\{0.7,0.7,0\}$, and $\{1,0.45,0.5\}$ for the outer, inner, and single loops, respectively. Disorder values used (from left to right): $V=\{0,5,12\}, W=\{0,1.2,1.6,2.0\}$. Calculations used $L=40$ for $\gamma$ and ${Ch}$,and $L=$ 800 for $E_g^\mathrm{min}$. Adapted from Ref. \cite{YPWu2022a}.
	}\label{fig12}
\end{figure*}

\subsubsection{Model and topology} 

A disorder-induced quantized pump was proposed in an optical superlattice of spinless fermions, which can be viewed as a dynamical version of TAIs and termed as topological Anderson–Thouless pump (TATP) \cite{YPWu2022a}.
The optical lattice system includes a dimerized hopping term \cite{Lohse2015a,Nakajima2021a}, a staggered on-site potential, and a Hubbard interaction term, which is described by the many-body Hamiltonian ${H}={H}_0+U\sum_j{n}_j{n}_{j+1}$. Here, the single-particle part with quasiperiodic modulations on the hopping and on-site potential reads \cite{YPWu2022a}.
\begin{equation}\label{YPWu2022a_H}	{H}_0=-\sum_j\left(J_j{c}_j^\dagger{c}_{j+1}+\mathrm{H.c.}\right)+\sum_j\left[(-1)^j\Delta+V_j\right]{n}_j,
\end{equation} 
where ${n}_j={c}_j^\dagger{c}_j$ is the number operator. The hopping amplitudes are dimerized: $J_j=J+\delta+W_j$ for even sites and $J_j=J-\delta$ and for odd sites, with $J\equiv1$ as the energy scale ( $\hbar/J\equiv1$ the unit of time). The quasiperiodic disorders are introduced in the on-site term $V_j=V\cos(2\pi\alpha j+\beta)$ and the hopping term $W_j=W\cos(2\pi\alpha j+\beta)$, where $V$ and $W$ represent the disorder strengths, $\alpha=(\sqrt5-1)/2$ is an irrational number ensuring incommensurability, and $\beta$ is an arbitrary phase. Both types of disorder have been experimentally realized in optical lattices \cite{Nakajima2021a,Meier2018a}. Here we focus on the noninteracting case of $U=0$ at the half-filling with particle number $N=L/2$ and system size $L$.

A Thouless pumping cycle is implemented by adiabatically varying the parameters on the $\Delta$-$\delta$ plane \cite{Thouless1983a}. The time-dependent pump protocol is described by the cyclic parameter $\phi(t) = 2\pi t/T \in [0, 2\pi]$, with period $T$. The potentials are modulated as $\Delta(t) = R_\Delta \sin \phi(t)$ and $\delta(t) = R_\delta \cos \phi(t) + \delta_s$. The slow evolution of $\phi(t)$ results in a quantized charge transfer per cycle, given by
\begin{equation}\label{YPWu2022a_DeltaQ}
	\Delta Q=Q(T)-Q(0)=\int_0^Tdt\:\langle\Psi(t)|{\mathcal{J}}(t)|\Psi(t)\rangle,
\end{equation}
where the averaged current operator is defined as ${\mathcal{J}}=\frac{i}{L}\sum_{j=1}^LJ_j{c}_j^\dagger{c}_{j+1}+\mathrm{H.c.}$.
In the presence of disorder or interactions \cite{QNiu1984a}, twisted PBC can be applied by introducing a phase twist $\theta \in [0, 2\pi]$ via the substitution ${c}_{j}^{\dagger}{c}_{j+1} \rightarrow e^{i\theta/L}{c}_{j}^{\dagger}{c}_{j+1}$ in the Hamiltonian (\ref{YPWu2022a_H}), denoted hereafter as ${H}_{\theta}$. The topology of the adiabatic pump can be characterized by the Chern number ${Ch}$ of the ground state $|\Psi_{g}\rangle$ of ${H}_{\theta}$, which is defined on the torus spanned by $(\theta, \phi)$ or equivalently $(\theta,t)$. The Chern number ${Ch}$ as the winding of the Berry phase $\gamma$ is given by ${Ch}=\int_0^{2\pi}\frac{d\phi}{2\pi}\partial_\phi\gamma=\int_0^T\frac{dt}{2\pi}\partial_t\gamma$.
In the adiabatic and thermodynamic limits, the averaged current satisfies \cite{QNiu1984a}
$\langle{\mathcal{J}}(t)\rangle=\int_0^{2\pi}\frac{d\theta}{2\pi}\langle\Psi_g(t)|\partial_\theta{H}_\theta|\Psi_g(t)\rangle=\frac{1}{2\pi}\partial_t\gamma$,
which implies the quantized charge pumping $\Delta Q = {Ch}$.

\subsubsection{Disorder-induced topological pumps} 

The Hamiltonian ${H}_{0}$ in Eq. (\ref{YPWu2022a_H}) reduces to the Rice-Mele model \cite{Rice1982a} in the absence of disorder, featuring a gap-closing point at $\Delta=\delta=0$ that constitutes a singularity in the Berry phase. A quantized pump necessitates an adiabatic parameter loop $\phi(t)$ that encircles this singular point without closing the energy gap, i.e., ensuring $E_{g}^{\min}=\min_{\phi}[E_{N+1}(\phi)-E_{N}(\phi)]>0$. Such a quantized pump can be destroyed by disorder-induced breaking of adiabaticity \cite{Rice1982a,Nakajima2021a,Hayward2021a}. Here, we reveal two different mechanisms for the disorder-induced quantized pumps \cite{YPWu2022a}, which are termed as extrinsic and intrinsic topological pumps driven from the on-site and hopping disorders, as shown in Figs.~\ref{fig12}(a) and \ref{fig12}(b), respectively.

We first analyze a specific pumping protocol comprising two nested loops of opposite orientation under on-site disorders, as illustrated in Fig.~\ref{fig12}(a). In the regime of weak or vanishing disorder $0\leqslant V\leqslant V_{c1}$, the energy gap closes exclusively at the point $(\Delta_c,\delta_c)=(0,0)$, with Chern numbers of the outer (inner) loops becoming ${Ch}_{\mathrm{out}}=1$ (${Ch}_{\mathrm{in}}=-1$). Since the total Chern number ${Ch}_\mathrm{total}={Ch}_\mathrm{out}+{Ch}_\mathrm{in}=0$, no net charge is pumped over a full cycle. As the disorder strength exceeds $V_{c1}$, the gapless region expands into a line along the $\Delta$ axis, widening progressively with increasing $V$. For two appropriately separated loops under intermediate disorder strengths $V_{c1} < V < V_{c2}$, the inner loop becomes trivial due to gap closure ($E_{g,\mathrm{in}}^{\min} = 0$), with the Chern number of outer loop retaining ${Ch}_\mathrm{out} = 1$. This enables an extrinsically induced topological pump with a net Chern number ${Ch}_\mathrm{total} = 1$, as shown in Figs.~\ref{fig12}(a) and \ref{fig12}(c). For stronger disorder $V > V_{c2}$, the gapless line intersects the outer loop, resulting in the gap closing and the breakdown of quantized pumps. One can numerically simulate the pump by computing the time evolution of the wave function
$|\Psi(t)\rangle={\mathcal{T}}\exp\left[-i\int_0^t{H}_0(t')dt'\right]|\Psi(0)\rangle$.
Here ${\mathcal{T}}$ is the time-ordering operator and the initial state $|\Psi(0)\rangle$ is constructed from the occupied eigenstates of ${H}_0(t=0)$ at half-filling. The pumped charges for the outer and inner loops, $\Delta Q_{\mathrm{out}}$ and $\Delta Q_{\mathrm{in}}$ are evaluated separately using Eq. (\ref{YPWu2022a_DeltaQ}), yielding the total charge transfer $\Delta Q_{\mathrm{total}}=\Delta Q_{\mathrm{out}}+\Delta Q_{\mathrm{in}}$. Numerical results for a system of size $L=80$ and pump period $T=400$ in Fig.~\ref{fig12}(c) demonstrate nearly quantized pump induced by moderate on-site disorder strength $V$.

We further propose an intrinsically disorder-driven topological pump, where quantized transport arises from a trivial single-loop pump by increasing hopping disorder strength $W$. The pump loop shown in Fig.~\ref{fig12}(b) initially possesses a Chern number ${Ch} = 0$ in the clean limit, without circling the gapless point at $(\Delta, \delta) = (0, 0)$. As $W$ increases, the critical point shifts to $\delta_c > 0$ and enters the pump cycle when $W_{c1} < W < W_l$. In the regime, the pump topology becomes nontrivial with ${Ch} = 1$. When $W_l<W<W_{c2} $, the gapless region evolves into a line that does not intersect the cycle, thereby preserving both the energy gap and nontrivial pump topology. The variations of the Chern number ${Ch}$ and the minimal gap $E_g^{\text{min}}$ with $W$ are displayed in Fig.~\ref{fig12}(d), confirming the emergence of a topological pump at intermediate hopping disorder strength $W_{c1}<W<W_{c2}$. In the disorder regime, the transferred charge $\Delta Q$ is nearly quantized to $\Delta Q \approx 1$, in contrast to the trivial pump with $\Delta Q \approx 0$ for $W = 0$, as shown in Fig.~\ref{fig12}(e) for systems of sizes $L = \{40, 80\}$ and pumping periods $T = \{200, 400\}$. Full quantization is expected in the thermodynamic limit \cite{YPWu2022a}. This quantized single-loop pump driven from a trivial pump by disorders provides a dynamical counterpart of TAI, and thus is termed as the TATP \cite{YPWu2022a}. The TATP breaks down when $W > W_{c2}$ (or when the gapless point exits the cycle) as the gapless line intersects the loop. The physical mechanism enabling the TATP is the disorder-induced shift of the gapless critical point. This effect can be further analyzed by the SCBA method, where disorder is incorporated by a self-energy $\Sigma_x(W)$. In this case, the quasiperiodic disorder yields the renormalized hopping parameter $\tilde{\delta}(W) = \delta + \Sigma_x(W)$, and the modified gapless point at $\delta_c =\tilde{\delta}(W) = 0$ \cite{YPWu2022a}. The shift of the gapless point based on the SCBA analysis is consistent with the numerical result, as shown in Fig.~\ref{fig12}(f).

The TATP is further shown to be generic \cite{YPWu2022a}: It appears in the  dynamic disorder and many-body interacting cases, and can be extended to 2D higher-order topological systems with disorder-induced quantized corner-to-corner transport. The effects of random on-site disorder on the topology and localization in higher-order Thouless pumping of noninteracting fermions were extensively explored in Ref. \cite{CWLu2023a}. The disorder response of the half-quantized topological Thouless pumping in a 1D system with the chiral anomaly was studied in Ref. \cite{ZQin2023a}. A disorder-induced topology of postquench states characterized by the quantized dynamical Chern number and the crossings in the entanglement spectrum in (1+1) dimensions was also predicted \cite{HCHsu2021a}. Recently, it was found that using the conventional single-loop protocol, a quantized Thouless pump can still be induced by on-site quasiperiodic disorders with staggered patterns \cite{Padhan2024a}. Moreover, the TATP was generalized to the non-Abelian regime, where a non-Abelian Thouless pump can be induced by quasiperiodic disorders in a Lieb chain with degenerate flat bands \cite{SHuang2024a}.

\subsection{Experimental realization} 

\begin{figure}[tb]
	\centering
	\includegraphics[width=0.48\textwidth]{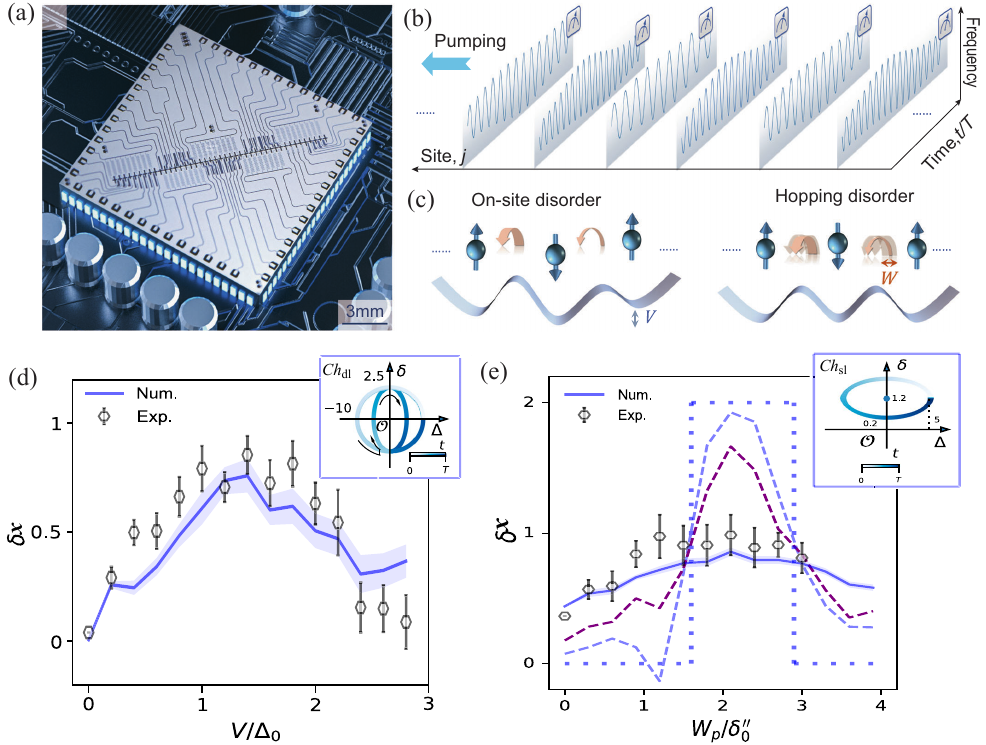}
	\caption{(Color online) (a) Optical micrograph of the 43-qubit superconducting chip. (b) Schematic of Floquet engineering with pulse sequences for the adiabatic cyclic evolution. The amplitude and the center shift of the pulse on each qubit are subject to a cyclic modulation, which correspond to the cyclic variations of hopping couplings and on-site potentials, respectively. (c) Schematic diagrams of the system with on-site staggered potential disorder (left) and hopping disorder (right).
	(d) Measured displacement of the center-of-mass $\delta x$ versus the on-site disorder strength $V$ during a double-loop pumping trajectory ${Ch}_{\mathrm{dl}}$ as shown in the inset.
	(e) Measured $\delta x$ versus the hopping disorder strength $W_{p}$ during a single-loop pumping trajectory ${Ch}_{\mathrm{sl}}$ as shown in the inset. The solid blue curve denotes the numerical simulation and the dotted curve shows the topological index calculated in the thermodynamic limit. The purple (blue) dashed curve denotes the numerical results using experimental parameters but with a longer period in a larger system. Adapted from Ref. \cite{YLiu2025a}.
	}\label{fig13}
\end{figure}

The two types of topological pumps induced by on-site potential disorder and hopping disorder, as proposed in Ref. \cite{YPWu2022a}, have been experimentally observed on a superconducting quantum simulator \cite{YLiu2025a}. The experiment was performed with 41 qubits on a tunable 1D superconducting processor, which consists of 43 nearest-neighbor-coupled transmon qubits, as shown in Fig.~\ref{fig13}(a). For single excitation, the tight-binding model Hamiltonian with on-site potential disorder or hopping disorder was experimentally simulated for pumping process \cite{YLiu2025a}: 
\begin{equation}\label{YLiu2025a_H}	\begin{aligned}{H}(t)&=\sum_{j=1}^{40}\Big\{J+(-1)^{j-1}\Big[\delta(t)+W_{j}\Big]\Big\}({a}_{j}^{\dagger}{a}_{j+1}+\mathrm{H.c.})\\&+\sum_{j=1}^{41}(-1)^{j-1}\Big[\Delta(t)+V_{j}\Big]{n}_{j}.\end{aligned}
\end{equation}
Here, $\alpha^\dagger$ and $\alpha$ denotes the hard-core bosonic creation and annihilation operator, respectively. $J\pm[\delta(t)+W_{j}]$ denote the hopping strengths with disorder $W_{j}$, $\pm[\Delta(t)+V_j]$ denote the staggered on-site potential with disorder $V_j$, and $\Delta(t)$ and $\delta(t)$ are periodic in time $t$ with the period $T$. The Floquet engineering technique was used to change the dynamical parameters $\delta(t)$ and $\Delta(t)$ adiabatically during a closed trajectory in a $\delta$-$\Delta$ space, as shown in Fig.~\ref{fig13}(b). Two types of disorders were also carefully introduced in the pumping process with the cyclic modulations of both the amplitude and the center offset of the engineering waves, corresponding to the cyclic variations of the hopping coupling and the on-site potential, respectively.

When the system is prepared as a Wannier state, the displacement of the center-of-mass per pumping cycle $\delta x={Ch}\cdot d$, where ${Ch}$ is Chern number of the pump and $d=2$ is the lattice constant. In experiment, the initial state is prepared as a single-excitation state by exciting one qubit closest to the middle, which has an overlap of over 0.99 with the Wannier state. During the pumping progress, the population of each qubit is measured to obtain the center-of-mass displacement after a pump cycle. The double-loop pumping procedure was first implemented to show the topological pumping induced by random on-site disorder, with a trajectory for opposite directions along the outer and inner loops [see Fig.~\ref{fig13}(d)]. In the clean case, there is no net pumped charge with zero total Chern number. As the on-site disorder strength $V$ increases, the gapless regime appears around the origin along the $\Delta$ axis. In a moderate disorder regime, the inner loop becomes trivial, while the outer loop remains nontrivial. The topological pumping induced by the on-site disorder was thus observed with $\delta x\neq0$, as shown in Fig.~\ref{fig13}(d). The second type of topological pumping induced by quasi-periodic hopping disorder was realized using a single-loop pumping trajectory \cite{YLiu2025a}. The single-loop pump trajectory is trivial as its center is biased away from the gapless point. As the disorder strength $W_p$ increases, the gapless point moves inside the pump loop. This leads to the nontrivial pumping, observed from the center-of-mass displacement $\delta x$, as shown in Fig.~\ref{fig13}(e).

\subsection{Other Floquet systems}
Floquet systems with periodical driving provide a dynamic platform for realizing disorder-induced topology. The trivial-to-topological transition in 2D driven systems indicates the emergence of Floquet TAIs  \cite{Titum2015a,Titum2017a}, which have been experimentally realized in a 2D photonic lattice \cite{Stuetzer2018a}, as introduced in Sec. \ref{SecIIC2}. The Floquet TAIs have recently been extended to higher-order topological systems \cite{Ghosh2023a}. The Floquet topological phase transitions induced by uncorrelated and correlated disorders have been examined in Ref. \cite{JHZheng2024a}.

By adding spatial disorders at an intermediate strength in 2D periodically driven systems, a new topological phase called the anomalous Floquet-Anderson insulator (AFAI) was predicted in Ref. \cite{Titum2016a}. The AFAI is characterized by a quasienergy spectrum featuring chiral edge modes and a fully localized bulk. This leads to a non-adiabatic quantized charge pumping even when all the bulk states are localized, which is absent in equilibrium systems. The topology of the AFAI is characterized by a topological invariant of a 3D frequency lattice with two spatial and one synthetic dimension \cite{Celi2014}. The non-equilibrium topological characters of the AFAI can also be revealed from quantized magnetization density \cite{Nathan2017a}. The AFAI with quantized response remains stable even in the presence of temporal noises, which break the time-periodicity of Floquet systems \cite{Timms2021a,PPZheng2023a}. A scheme to realize the AFAI with a continuously driven optical lattice was proposed in Ref. \cite{Dutta2024a}.
Recently, the disorder-driven topological phase transitions via topological edge modes have been experimentally observed with ultracold atoms in periodically-driven 2D optical lattices \cite{Hesse2025}. This work confirms that disorder favors the Floquet topological regime and paves the way towards realizing the non-equilibrium AFAI.

Another kind of adiabatic quantized pump that has no analogy in undriven systems is the topological Floquet-Thouless energy pump \cite{Kolodrubetz2018a}, where the quantized pumped quantity is energy rather than charge. This phenomenon is stabilized by disorder and topologically characterized by the 3D winding number of the micromotion with respect to time within each cycle, momentum, and adiabatic tuning parameter. Recently, the Floquet-Thouless energy pump was extended to quasiperiodically driven systems \cite{Nathan2021a,Long2021a}, where the time-translation symmetry is broken and disorders play an important role. For a disordered 1D fermionic system under quasiperiodic driving with two modes of incommensurate frequencies, the non-adiabatic quantized energy pump is protected by a combination of disorder-induced spatial localization and frequency localization, which is unique to quasiperiodically driven systems.

\section{Disorder-induced many-body topological phases}

We now review recent theoretical and experimental advances on disorder-induced interacting topological phases in many-body fermionic and bosonic systems.

\subsection{Fermionic systems}

\subsubsection{One-dimensional model}
\begin{figure}[tb]
	\centering
	\includegraphics[width=0.46\textwidth]{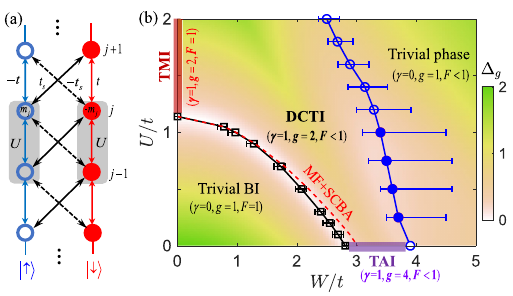}
	\caption{(Color online) (a) Schematic of the model Hamiltonian described by Eq. (\ref{TAMI-Ham}). (b) Phase diagram on the $W$-$U$ plane. The square and circle symbols guided by solid lines are obtained from the averaged Berry phase $\gamma\approx0.5$, and the error bars denote the uncertainty region with non-quantized values of $\gamma$. The red dashed line denotes the phase boundary between the trivial band insulator (BI) and the topological phases, determined from the mean-field (MF) and SCBA approaches. The background colormap represents the excitation gap $\Delta_g$. Parameters are $t=1$, $t_s=0.95$, $m_z=3$, $L=N_{a}=24$ for the four solid blue circles from the DMRG, otherwise $L=N_a=8$ for $U\neq0$ and $L=610$ for $U=0$ from the exact diagonalization. Adapted from Ref. \cite{GQZhang2021a}.
	}\label{fig14}
\end{figure}

We consider spin-1/2 fermionic atoms with repulsive interactions confined to a 1D spin-dependent optical lattice, which incorporates disordered ladder potentials \cite{Creutz1999,Atala2014}, as shown in Fig.~\ref{fig14}(a). The system is governed by the following many-body Hamiltonian \cite{GQZhang2021a}: 
\begin{equation}\label{TAMI-Ham} 
\begin{aligned}
	H=&-t\sum_j(c_{j,\uparrow}^\dagger c_{j+1,\uparrow}-c_{j,\downarrow}^\dagger c_{j+1,\downarrow}+\mathrm{H.c.})\\&+t_s\sum_j(c_{j,\uparrow}^\dagger c_{j+1,\downarrow}-c_{j,\downarrow}^\dagger c_{j+1,\uparrow}+\mathrm{H.c.})\\&+\sum_jm_j(n_{j,\uparrow}-n_{j,\downarrow})+U\sum_{j,\sigma=\uparrow,\downarrow}n_{j,\sigma}n_{j+1,\sigma},
\end{aligned}
\end{equation}
where $n_{j,\sigma}=c_{j,\sigma}^\dagger c_{j,\sigma}$ denotes the particle number operator, and the creation operator $c_{j,\sigma}^\dagger$ creates a fermion with spin $\sigma$ ($\sigma = \uparrow, \downarrow$) at lattice site $j$ ($j=1,2,\ldots,L$). Here, $t$ (let $t\equiv1$ as the energy unit) and $t_s$ represent the spin-conserving and spin-flip hopping amplitudes, respectively, and $U$ parameterizes the intraleg interaction strength. The term $m_j = m_z + W\cos(2\pi\alpha j + \varphi)$ denotes a Zeeman potential with quasiperiodic disorders, where $m_z$ denotes a uniform Zeeman strength, $W$ controls the disorder strength, $\alpha=(\sqrt{5}-1)/2$ is set as the irrational number to ensure incommensurability, and $\varphi$ is a random phase offset for sample averaging. We focus on the half-filling case with the total atom number $N_a=L$, with $L$ being the lattice length. The required optical lattice can be realized using proper laser beams with incommensurate modulations for two atomic internal states of opposite detunings. The spin-flip and spin-dependent hoppings have been experimentally realized by the Raman-assisted tunneling technique \cite{ZWu2016,BSong2018}. The interaction terms can be engineered for dipolar atoms with long-range intra-leg interactions \cite{Juenemann2017}. 

In the clean and noninteracting limit, the model supports a single-particle topological insulating phase protected by the chiral symmetry when $|m_z| < 2t$ \cite{XJLiu2013}. This chiral symmetry preserves in the presences of disorders and interactions. The many-body ground states of Hamiltonian (\ref{TAMI-Ham}) are numerically studied by using the exact diagonalization and density-matrix renormalization group methods \cite{Schollwoeck2005}, and further analyzed using the mean-field and SCBA approaches.

\subsubsection{Phase diagram}

The topological nature of this 1D interacting system can be captured by a many-body Berry phase quantized in units of $\pi$ under twisted PBCs \cite{Juenemann2017}:
\begin{equation}
	\gamma=\frac1\pi\oint d\theta\left\langle\Psi_g(\theta)\right|i\partial_\theta\left|\Psi_g(\theta)\right\rangle\mathrm{mod}~2,
\end{equation} 
where $\left| \Psi_g(\theta) \right\rangle$ denotes the half-filled many-body ground state in the presence of a twist angle $\theta$. The value $\gamma = 1$ corresponds to a topological phase, while $\gamma = 0$ for a trivial phase. 
For interacting topological phases, their ground states under OBCs near half-filling are twofold degenerate with each collective excitation occupied on each edge. The ground state degeneracy $g=2$ counts the degeneracy among energies $\{E_0^L,E_1^L,E_0^{L\mp1}\}$ with $E_0^{L}=E_1^{L}\neq E_0^{L\mp1}$, with $E_{0,1}^{N_\alpha}$ the two lowest energies for $N_a=L-1,L,L+1$ under OBC. For non-interacting TAI under OBCs, two zero-energy single-particle edge modes can either be empty or occupied, such that $E_0^{L}=E_1^{L}=E_0^{L\mp1}$ and the degeneracy $g=4$. For trivial phases, the degeneracy $g=1$ without edge modes. 
In addition, the overlap between different ground-state wave functions against disorders is characterized by the fidelity $\mathcal{F}=\frac2{N_d(N_d-1)}\sum_{i\neq j}\langle\Psi_g(\varphi_i)|\Psi_g(\varphi_j)\rangle$, where $\varphi_{i,j}$ are random phases from $N_{d}$ disorder realizations. The fidelity $\mathcal{F}\approx 1$ indicates an ordered phase, while $\mathcal{F}<1$ for a disordered phase.

The main results are summarized in the phase diagram in the $W$–$U$ parameter space, as shown in Fig.~\ref{fig14}(b). The background color denotes the value of the gap $\Delta_g=E_1-E_0$ between the ground and first excited states with energies $E_{0,1}$ under PBCs. The ordered and disordered phases are identified via the fidelity $\mathcal{F}$. The topological properties are characterized by the quantized Berry phase $\gamma$
and the corresponding ground state degeneracy $g$ under OBCs. For small $W$ and $U$, the system hosts a trivial band insulator with long-range order. By increasing $U$ ($W$) up to the topological phase transition point $U_c\approx 1.14$ ($W_{c}\approx 2.8$) in the clean limit $W=0$ (noninteracting limit $U=0$), the ground state becomes an ordered topological Mott insulator (TMI) \cite{Raghu2008a,Pesin2010a,YZhang2009a,Yoshida2014a,Herbut2014a,YLChen2020a} (disordered TAI). For intermediate values of $U$ and $W$, a disordered correlated topological insulator (DCTI) emerges, which can be driven from the trivial band insulator by increasing $W$ for $U \lesssim 1.14$. For strong disorder, both of the DCTI and TAI become a disordered trivial phase. One can see that the bulk gap closing and reopening in the first and second phase boundaries in the phase diagram, respectively. 
Thus, the TAI and TMI are adiabatically connected via the DCTI without gap closure. This intermediate phase provides a unified description for both disorder- and interaction-induced topological phases.
Moreover, the DCTI behaves as a disordered TMI when $U\gtrsim1.14$ and a correlated TAI in the other region, which generalizes the concept of TAIs to the many-body interaction regime. 
In the phase diagram, the first phase boundary between the trivial band insulator and topological phases are extracted from the Berry phase and other complementary quantities, which agree well with the following analysis.

By combining the mean-field and SCBA techniques, a unified framework has been developed to identify topological phase transition driven by interaction and disorder \cite{GQZhang2021a}. Under the Hartree-Fock approximation, the density–density interaction term in Hamiltonian (\ref{TAMI-Ham}) is linearized as $n_{j,\sigma} n_{j+1,\sigma}\approx\left\langle n_{j,\sigma}\right\rangle n_{j+1,\sigma}+\left\langle n_{j+1,\sigma}\right\rangle n_{j,\sigma}-\langle n_{j,\sigma}\rangle\left\langle n_{j+1,\sigma}\right\rangle$.
In the band insulator regime, it can be further simplified to $\sum_j n_{j,\sigma}n_{j+1,\sigma}\approx\rho_s\sum_j(n_{j,\uparrow}-n_{j,\downarrow})$ up to a constant, where $\rho_s$ denotes the spin density difference. Thus, the interaction effectively renormalizes the Zeeman term. By numerical evaluation of $\rho_s$, the renormalized Zeeman strength is $\tilde{m}_z \approx m_z - U/1.14$. The disorder effect can be incorporated from the self-energy term $\Sigma(W)$ based on the SCBA method. In this model, the term $\Sigma(W)$ obeys the self-consistent equation
\begin{equation}	\frac{1}{E_f-H_{\mathrm{MF}}(k)-\Sigma(W)}=\left\langle\frac{1}{E_f-H_{\mathrm{eff}}(k,W_q)}\right\rangle_q.
\end{equation}
Here, $E_f=0$ is the Fermi energy, $H_\mathrm{MF}(k)=(\tilde{m}_z-2t\cos k)\sigma_z-2t_s\sin k\sigma_y$ is the clean mean-field Hamiltonian, $H_\mathrm{eff}$ represents the effective Hamiltonian modified by
disorder $W_q=W\cos(2\pi\alpha q)$ with $q=1,2,\ldots,N_{q}$, and $\langle\cdots\rangle_q$ denotes averaging over $N_q$ disorder configurations.
By computing $\Sigma=\Sigma_{y}\sigma_y+\Sigma_{z}\sigma_z$, one can obtain the renormalized Zeeman strength $\bar{m}_{z}=\tilde{m}_{z}+\Sigma_{z}\approx m_{z}-U/1.14+\Sigma_{z}$ and the irrelevant renormalized hopping strength $\bar{t}_{s}=t_s+\Sigma_{y}$. The topological phase boundaries are determined by $|\bar{m}_{z}|=2$, as shown in Fig.~\ref{fig14}(b), agreeing well with the numerical results. Under strong disorder and interaction strengths, this approach does not apply to capture the second transition from the topological phases to the trivial phase. The effect of 1D commensurate moiré potentials on topological and correlated phases of spinful fermions was explored in Ref. \cite{GQZhang2025a}. A reentrant topological phase (with a sequence of trivial-to-topological transitions) from a trivial phase has been revealed, which shares similar renormalization mechanism with the noninteracting and interacting TAIs.

\subsubsection{Two dimensions}
The disorder effect on interacting topological insulators of 2D fermions has been investigated in Ref. \cite{HHHung2016a}. Although the absence of  disorder-induced TAIs, it was found that the topological phases are stable against disorders even when interactions are included. In the spinful Harper-Hofstadter-Hatsugai model on a 2D square lattice \cite{Harper1955,Hofstadter1976a,Hatsugai1990}, it was found that the repulsive on-site interaction can broaden the regime of the topological phase under weak disorders \cite{JHZheng2019a}. Recently, the ground-state phase diagram of spinless fermions in a 2D Haldane-Hubbard model under on-site random disorders has been investigated in Ref. \cite{TCYi2021a}. It was found that the interplay of disorder, interactions, and topology gives rise to a rich phase diagram. In particular, an interacting TAI that is driven by disorder from a trivial phase under finite interactions was revealed in this 2D system. For similar 2D disordered Haldane model with extended Hubbard interactions, a mean-field calculation has been performed to study the disorder-driven interacting TAIs and related phase transitions \cite{Silva2024a}. In addition, an exactly solvable model of 2D disordered topological superconductor with Hubbard interactions was studied and the interacting TAI was found in this system \cite{YTDeng2024a}.

\begin{figure}[tb]
	\centering
	\includegraphics[width=0.46\textwidth]{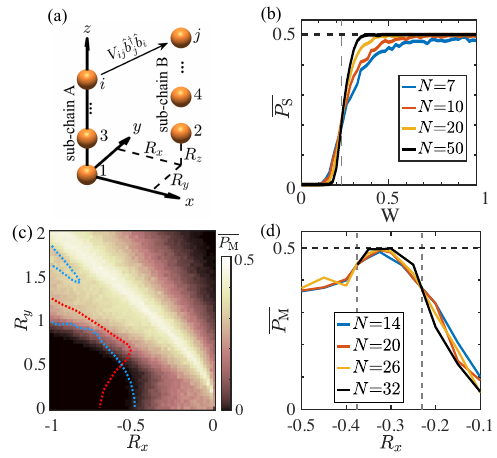}
	\caption{(a) Schematics of a dimerized Rydberg atom chain described by the Hamiltonian in Eq. (\ref{RydHam1}). Two subchains with atoms $2i-1$ and $2i$ form a unit cell, whose position vector is $\boldsymbol{R}=(R_x,R_y,R_z)$. (b) Polarization $\overline{P_{S}}$ for single-particle case versus the disorder strength $W$ when $R_x=-0.5$ and $R_y=0$. (c) Topological invariant $\overline{P_{M}}$ for many-body case versus $R_x$ and $R_y$ for an amorphous lattice with $N=8$. Cyan dotted lines 
and red dotted line are phase boundaries for a regular lattice $(N=8)$ of hard-core bosons at half-filling and for an amorphous lattice $(N=200)$ in the single-particle level, respectively. (d) $\overline{P_{M}}$ versus $R_x$ for various system sizes with $R_y=1$. Adapted from Ref. \cite{KLi2021a}.
	}\label{fig15}
\end{figure}

\begin{figure}[tb]
	\centering
	\includegraphics[width=0.46\textwidth]{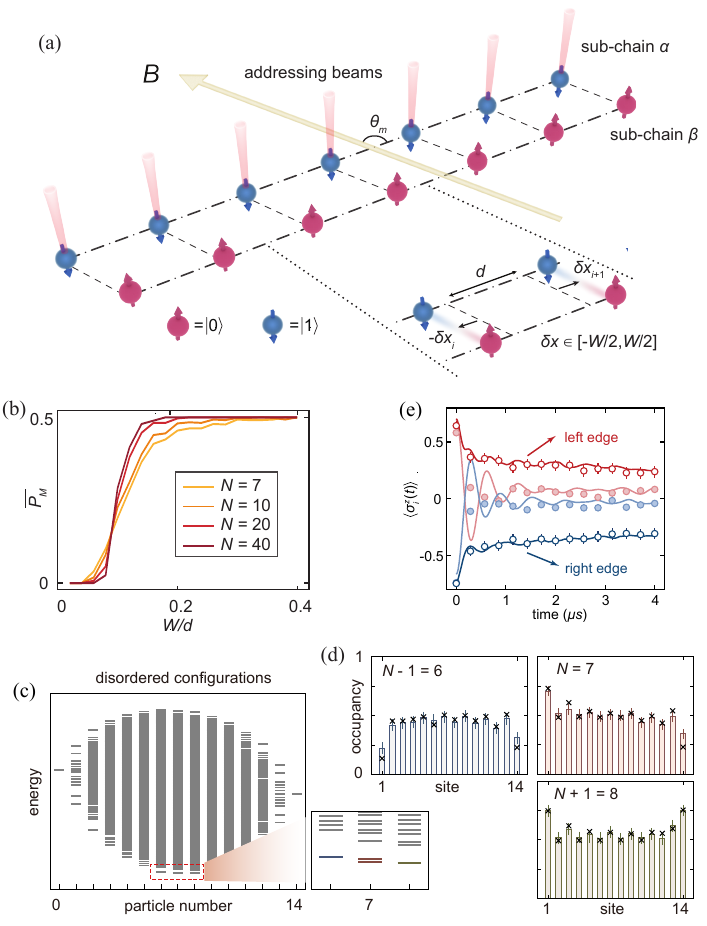} 
	\caption{(a) Schematics of a 1D dimerized Rydberg atom chain with two sub-chains and structural disorder, described by the hard-core Hamiltonian in Eq. (\ref{RydHam2}). The light red tubes denote the addressing laser beams used to prepare a N\'{e}el state, and $B$ denotes the magnetic field. (b) Numerically calculated average topological invariant $P_M$ as a function of the structural disorder strength $W$, showing a topological phase transition from a trivial state at $\overline{P_{M}}=0$ to a nontrivial phase at $\overline{P_{M}}= 0.5$. 
(c) Calculated Energy spectra at different particle numbers for a disordered lattice $(W=0.4d)$. A zoomed-in view shows the four-fold degeneracy of the ground state, which is slightly lifted due to the van der Waals interactions. (d) Measured site occupancy of the ground state for the disordered lattice near half-fillings of $N-1$ (left), $N$ (right), and $N+1$ (below) particles. 
(e) Measured magnetization $\sigma_i^z(t)$ of two edge sites from the initial N\'{e}el state, with empty and filled circles for the disordered and regular lattices, respectively. Solid lines denote the numerical predictions. Adapted from Ref. \cite{ZPYue2025a}.
	}\label{fig16}
\end{figure}

\subsection{Bosonic systems}

\subsubsection{Proposal with Rydberg atom arrays}

In Ref. \cite{KLi2021a}, the authors studied topological phases in a 1D amorphous Rydberg atom chain with random atom configurations. In the single-particle level, the topological amorphous insulators induced by the structural disorder was found. Moreover, a symmetry-protected topological phase of interacting bosons in this disordered system was predicted. The proposed system is a dimerized Rydberg atom chain with $2N$ atoms, as shown in Fig.~\ref{fig15}(a). Each atom has two Rydberg states, denoted by $|s\rangle=|0\rangle$ and $|p\rangle={b}^\dagger|0\rangle$. The transitions between these states are induced by the dipolar interaction between atoms, and the system can be captured by the Hamiltonian \cite{KLi2021a,Leseleuc2019a}:
\begin{equation}\label{RydHam1}	{H}=\sum_{i<j}^{2N}J_{ij}({b}_i^\dagger{b}_j+{b}_j^\dagger{b}_i).
\end{equation}
Here, ${b}_{i}^{\dagger}$ (${b}_{i}$) is creation (annihilation) operator for a hard-core boson at site $i$. The hopping amplitude $J_{ij}=d^{2}(1-3\cos^{2}\theta_{ij})/R_{ij}^{3}$ arises from the dipolar interaction, where $d$ is the atomic dipole moment, and $\theta_{ij}$ denotes the angle between the external magnetic field $\boldsymbol{B}$ and the interatomic vector $\boldsymbol{R}_{ij}$. In the dimerized chain, each cell contains one atom from sublattice A (odd sites) and one from sublattice B (even sites), which are separated by a vector $\boldsymbol{R} = (R_x, R_y, R_z)$. The intracell hopping is given by $J' = d^2 (R_y^2 - R_x^2 + 2\sqrt{2}R_xR_z)/R^5$. By aligning the chain at the magic angle $\theta_m = \arccos(1/\sqrt{3})$ relative to the magnetic field, the hopping within each sublattice is suppressed and the chiral symmetry is achieved. The structural disorder is introduced by displacing atoms from their ordered positions via $z_{2i-1} \to i-1 + \delta z_i$ and $z_{2i} \to i-1 + R_z + \delta z_i$, with each $\delta z_i$ being uniformly drawn from $[-W/2, W/2]$ and disorder strength $W$.

When there is only one excitation in the Rydberg atom chain, the system is governed by a single-particle Hamiltonian $H^S$, with matrix elements $[H^S]_{ij} = J_{ij}(1 - \delta_{ij})$ for $1 \leq i, j \leq 2N$, in the basis $\beta = \{{b}_1^\dagger|0\rangle, \ldots, {b}_{2N}^\dagger|0\rangle\}$. As $H^{S}$ has chiral symmetry: $\Gamma H^{S} \Gamma^{-1} = -H^{S}$, with $\Gamma = \text{diag}\{(-1)^{j-1}\}_{j=1}^{2N}$, the topological property can be characterized by the polarization \cite{Resta1998a}
\begin{equation}
	P_S=\left(\frac{1}{2\pi}\text{Im}\{ \ln [\det (U^\dagger DU)]\} -\frac{1}{2N}\sum_{i=1}^{2N}x_i\right) {\mathrm{mod}} ~1.
\end{equation}
Here the columns of $U$ are the negative-energy eigenstates of $H^{S}$ under PBCs, and $D = \mathrm{diag}\{e^{2\pi i x_{j}/N}\}_{j=1}^{2N}$ with $x_{i}$ being the atomic positions. Due to chiral symmetry, $P_{S}$ quantizes to 0 and 0.5 for the trivial and nontrivial phases, respectively. Figure \ref{fig15}(b) shows the configuration-average $\overline{P_S}$ versus disorder strength $W$ for various system sizes. One can find a transition from trivial to topological phase at $W \approx 0.23$, which indicates the TAI induced by structural disorders.

In the multi-excitation regime, the system becomes a true many-body system with interactions because the Hamiltonian (\ref{RydHam1}) with long-range hopping cannot be mapped to a free fermionic Hamiltonian. In this case, the many-body topological invariant $P_M$, analogous to $P_S$ for the single-particle case, can be defined as  
\begin{equation}\label{PM}
	P_M = \frac{1}{2\pi} \mathrm{Im} \left( \ln \langle \Psi_g | {\mathcal{P}}_M | \Psi_g \rangle  \right) .
\end{equation}
Here, $| \Psi_g \rangle$ is the many-body ground state of the hard-core bosonic Hamiltonian for periodic boundaries, and ${\mathcal{P}}_M = \prod_{j=1}^{2N} e^{-(i\pi /N) x_j \sigma_j^z}$ is the twist operator.

The phase diagram in Fig.~\ref{fig15}(c) shows a long narrow region with $\overline{P_{M}}\approx0.5$, which indicates the existence of an interacting topological phase for amorphous hard-core bosons at half-filling. For comparison, the phase boundaries for the single-particle case in amorphous lattices and the many-body case in regular lattices are plotted as dotted lines. Notably, there exists a region around $R_y=1.75$ where the phase is trivial for a regular lattice while the phase is in an intermediate region with $0<\overline{P_{M}}<0.5$ for an amorphous lattice. This signals a possible interacting TAI phase in the many-body case induced by structural disorders, although the related topological phase transition is still unclear due to the system size limitation. To further identify the existence of the many-body topological phase in the amorphous system, $\overline{P_{M}}$ versus $R_{x}$ for various system sizes are shown in Fig.~\ref{fig15}(d). A parameter region for $-0.375\lesssim R_x\lesssim-0.23$ (other regions such as $R_{x}>-0.23$) with $\overline P_M\approx0.5$ is approaching (declining) as $N$ increases, which suggests a topological (trivial) phase.

In a recent work \cite{PHe2025a}, amorphous topological phases for arrays of randomly pointed Rydberg atoms have been extended to nonequilibrium regime by periodic driving. At the single-particle level, structural disorder induces Floquet TAI with localized bulk states and two types of ($0$-type and $\pi$-type) edge modes. The Floquet amorphous topological order still exists in the many-body case, and is captured by the topological entanglement entropy and the string order. The interplay of many-body localization and topology in disordered dimerized array of Rydberg atoms has been investigated in Ref. \cite{Prodius2025a}. It was found that the topological states exist across the entire energy spectrum with localized region characterized by Hilbert space fragmentation.

\subsubsection{Experimental realization}

In a recent experiment \cite{ZPYue2025a}, the disorder-induced interacting average symmetry protected topological phase has been realized in a Rydberg atom array. Similar to the proposal in Ref. \cite{KLi2021a}, the Rydberg atom array is dimerized with two sub-chains (denoted by $\alpha$ and $\beta$), and the structural disorder is introduced by randomly displacing each unit cell
from their regular lattice sites along the sub-chain direction, as shown in Fig.~\ref{fig16}(a). The dipolar exchange interaction induces the flip-flop hoppings between atoms $i$ and $j$. In the experiment, two Rydberg atoms in the vacuum state also experience the van der Waals interaction. The system can be described by the 1D staggered hard-core boson model with structure disorder \cite{ZPYue2025a,KLi2021a}
\begin{equation} \label{RydHam2} {H}=\sum_{i<j}^{2N}J_{ij}({b}_{i}^{\dagger}{b}_{j}+{b}_{j}^{\dagger}{b}_{i})+ V_{ij}^{\mathrm{vdW}} (1-{n}_i)(1-{n}_j),
\end{equation}
with the vacuum and one-particle excitation states encoded into the Rydberg $|s\rangle=|0\rangle$ and $|p\rangle={b}^\dagger|0\rangle$ states, respectively. Here, ${b}_j$ $({b}_j^\dagger)$ is the annihilation (creation) operator of a boson at site $j$, constrained by the hard-core condition ${b}_j^2=({b}_j^\dagger)^2=0$, and ${n}_i={b}_i^\dagger{b}_i$ is the particle number operator. The random position vector is introduced in the hopping $J_{ij}$ as the structural disorder of strength $W$ \cite{KLi2021a}, which is realized via random displacements to the centers of tweezers from their regular lattice sites. The strength of the additional van der Waals interaction $V_{ij}^{\mathrm{vdW}}$ is comparable to $J_{ij}$ in the experiment, which contributes an Ising term of the form $\sum_{i<j}\sigma_i^z \sigma_j^z$. A nonzero $V_{ij}^{\mathrm{vdW}}$ breaks the sublattice (chiral) symmetry. However, all the Hamiltonians on random lattice configurations constitute an ensemble respecting an average inversion symmetry \cite{ZPYue2025a}, which can protect topological phases \cite{RCMa2023a,RCMa2025a}.

In experiment, a structural-disorder induced topological phase in the single-particle level was first observed using the microwave spectroscopy of the local density of states \cite{ZPYue2025a}. Then the many-body ground states with particle numbers near half-filling were prepared to reveal the disorder-induced interacting topological phase, which is characterized by the many-body topological invariant $P_{\mathrm{M}}$ in Eq. (\ref{PM}). Figure \ref{fig16}(b) shows that averaged over many independent realizations, a sharp transition from a trivial state with $\overline{P_{M}}=0$ when increasing structural disorder strength is observed, indicating a many-body average symmetry-protected topological phase with $\overline{P_{M}}=0.5$ \cite{RCMa2023a,RCMa2025a}. The induced topological many-body ground state was experimentally revealed from its degeneracy and the edge modes. The measured site-dependent occupancies of the prepared ground state in disordered lattices are shown in Fig.~\ref{fig16}(d). At half-filling, the disordered configuration has pronounced differences for either the left or right edges with occupation imbalance. Two statistically degenerate ground states exist in the half-filling subspace, as shown in Fig.~\ref{fig16}(c). Furthermore, a compelling evidence that the disorder-induced topological phase exhibits robust edge states was provided by measuring the time evolution of the magnetization $\sigma_i^z(t)$ of all sites from the initial N\'{e}el state. As shown in Fig.~\ref{fig16}(e),  
$\langle\sigma_{i}^{z}\rangle$ at the two edges exhibits a slower decay and eventually stabilizes at finite values in the disordered lattice, in contrast to the fast decay in the regular lattice. This is consistent with the presence of topologically protected edge modes in disordered lattices.

In another recent experiment \cite{LSu2025a}, the authors observed topological phase transitions and the emergence of mixed-state order using a quantum simulator of interacting erbium atoms in a 1D optical lattice. They realized a staggered dipolar Bose-Hubbard model, where a transition between two distinct crystalline symmetry-protected topological phases is identified through measuring non-local string orders. A key finding is that while introducing symmetry-breaking disorder destroys the sharp transition in any single realization, averaging over disorder restores it. This recovery highlights that the adjacent phases represent distinct mixed-state topological orders, stabilized by the restoration of inversion symmetry on average. This work provides the experimental demonstration of average symmetry-protected topological phases and illustrates that topological criticality can be restored in disordered ensembles. Notably, the mixed-state order has also been probed via Renyi correlators and variational decoding in a digital quantum device of trap ions under dephasing noise channels \cite{YZhang2025a}.

\section{Overview and outlook}

In summary, we have introduced the recent theoretical and experimental progress on disorder-induced topological phases in both condensed-matter and artificial systems. By means of concrete models with random and quasiperiodic disorders, we have illustrated various TAI phases with different localization properties of bulk states and topological Anderson transitions. We have summarized the non-Hermitian extensions of TAI phases with unique topological and localization characters under the combination of  disorders and non-Hermiticities. We have further reviewed the disorder-induced topology in dynamical and many-body systems, by highlighting topological Anderson-Thouless pumps, disordered correlated topological insulators and average-symmetry protected topological orders. We have also surveyed the growing experimental advances in realizations and detections of these disorder-induced topological phases.
{ Ultimately, these diverse phenomena are unified by the mechanism of disorder-driven renormalization. As elucidated through the framework of effective medium theory, the self-energy contribution from disorder effectively modifies the topological terms, allowing non-trivial phases to emerge from trivial ones. This common physical origin manifests through distinct pathways across different systems and regimes. In random systems, it is characterized by ensemble-averaged stability, while in quasiperiodic systems, it is defined by the formation of mobility edges. For Floquet or pump systems, the transition is driven by the disorder-induced shifting of gapless point. Finally, in non-Hermitian and interacting systems, this mechanism extends to point-gap and many-body  topologies. }

Finally, we would like to end by highlighting some possible directions for future studies. The effect of some unconventional types of disorders on topological phases remains to be explored, such as binary \cite{SNLiu2022a} and power-law correlated disorders \cite{Girschik2013a}, non-Hermitian disorders \cite{XWLuo2023a,CZChen2021a,QYMo2022a}, and disordered (synthetic) gauge fields (e.g., random magnetic flux). In systems like quantum Hall and ladder models \cite{Creutz1999,Juenemann2017}, a uniform magnetic flux is known to profoundly alter band topology. Introducing spatially random flux could lead to a novel disorder-induced topological transition sequence between different topological phases \cite{CALi2025a}. Whether random real or imaginary flux can drive a trivial insulator into a topological phases with different localization states is still an open question to be investigated. While there are a few initial experimental studies of TAIs in materials like HgTe and MnBi$_4$Te$_7$ \cite{Khudaiberdiev2025a,AQWang2025a},
more explorations of disorder-driven topological materials are required. Recent studies on chiral magnets like FeGe demonstrate that increasing disorder via ion irradiation can significantly expand the stable regime of topological phases, broadening the temperature range of topological Hall signals and modifying anomalous Hall scaling \cite{Gupta2025}. This highlights that disorder engineering would be a powerful tool for stabilizing topological textures and detecting quantized electrical response in condensed-matter materials. Preliminary investigations have revealed exotic disorder-induced topological phases in non-Hermitian disordered Hamiltonian systems \cite{DWZhang2020a,XWLuo2023a,LZTang2020a,Claes2021a,QLin2022a,BBWang2025a}.
An open question is the disorder-induced topology in the Liouvillian framework for open quantum systems. In particular, it be worth extending NHTAIs to the realm of open quantum systems. A promising direction lies in dissipation-engineered TAIs, where engineered gain-and-loss \cite{Bender1998a,LFeng2017a,ElGanainy2018a,Miri2019a,QYMo2022a} may stabilize topological states or even induce topological phases through purely dissipative disorders. 
{In addition, drawing from the burgeoning field of measurement-induced phase transitions  \cite{YDLi2018,Koh2023, HZLi2025a,XJYu2026,SZLi2024,XZFeng2025,Matsubara2025,Yokomizo2025} in monitored quantum circuits \cite{HZLi2025a}, it is interesting to investigate whether the random projective measurements can give rise to topological phases.}
Another direction worth exploring is about disorder-induced topological phenomena in many-body interacting systems in equilibrium and out-of-equilibrium regimes \cite{GQZhang2021a,KLi2021a,PHe2025a,YPWu2022a,ZPYue2025a,LSu2025a}. Recently,  the 1D many-body interacting SPT phases protected by disorder-average symmetry \cite{RCMa2023a,RCMa2025a} have been observed in tunable quantum simulators \cite{ZPYue2025a,LSu2025a}. The synthetic quantum systems offer a platform to simultaneously integrate and control disorders, interactions, and dissipation (non-Hermiticity). Thus, it is promising to
further examine disorder-induced many-body topological phases in higher dimensions, non-Hermitian regime, and Floquet case by incorporating periodic driving. Moreover, the interplay between many-body localization and disorder-induced topology in interacting systems remains an open challenge \cite{Hamazaki2019a,LJZhai2020a,Prodius2025a}. The fate of topological order in the many-body localization regime requires further theoretical and experimental investigations. In conclusion, there are many interesting aspects of disorder-induced topological phases remain to be explored, and the present review may help crystallize current and further studies in this field.

\begin{acknowledgments}
We appreciate the collaboration with S.-L. Zhu, H. Yan, G.-Q. Zhang, L.-J, Lang, Y.-P. Wu, S.-N. Liu, L.-F. Zhang, W.-J. Zhang, S. P. Zhao, H. Yu, J. Zhang, Y. Jin, and X. Li on the topic. This work is supported by the Guangdong Basic and Applied Basic Research Foundation (Grant No. 2024B1515020018), the National Natural Science Foundation of China (Grant No. 12174126), and the Science and Technology Program of Guangzhou (Grant No. 2024A04J3004). 
\end{acknowledgments}

\bibliography{reference}
\clearpage


\end{document}